\documentclass[twocolumn,showpacs,amsmath,amssymb,aps, prb,superscriptaddress]{revtex4-1}

\pdfoutput=1
\usepackage[]{graphicx}
\usepackage[]{verbatim}
\usepackage[]{color}
\usepackage{overpic}
\usepackage{rotating}
\usepackage{mathtools}
\usepackage{appendix}
\usepackage{ulem}
\usepackage[margin=0.6in]{geometry}

\makeatletter
\newsavebox{\@brx}

\newcommand{\llangle}[1][]{\savebox{\@brx}{\(\m@th{#1\langle}\)}%
  \mathopen{\copy\@brx\mkern2mu\kern-0.8\wd\@brx\usebox{\@brx}}}
\newcommand{\rrangle}[1][]{\savebox{\@brx}{\(\m@th{#1\rangle}\)}%
  \mathclose{\copy\@brx\mkern2mu\kern-0.8\wd\@brx\usebox{\@brx}}}

  \newcommand{\lllangle}[1][]{\savebox{\@brx}{\(\m@th{#1\langle}\)}%
  \mathopen{\copy\@brx\copy\@brx\mkern4mu\kern-0.7\wd\@brx\usebox{\@brx}}}
\newcommand{\rrrangle}[1][]{\savebox{\@brx}{\(\m@th{#1\rangle}\)}%
  \mathclose{\copy\@brx\copy\@brx\mkern4mu\kern-0.7\wd\@brx\usebox{\@brx}}}

\graphicspath{{./}{./}}

\begin{document}
\title{Path to stable quantum spin liquids in spin-orbit coupled correlated materials }
\author{Andrei Catuneanu}
\affiliation{Department of Physics and Center for Quantum Materials, University of Toronto, 60 St.~George St., Toronto, Ontario, M5S 1A7, Canada}
\author{Youhei Yamaji}
\affiliation{Department of Applied Physics and Quantum-Phase Electronics Center (QPEC), The University of Tokyo, Hongo, Bunkyo-ku, Tokyo, 113-8656, Japan}
\affiliation{JST, PRESTO, Hongo, Bunkyo-ku, Tokyo, 113-8656, Japan}
\author{Gideon Wachtel}
\affiliation{Department of Physics and Center for Quantum Materials, University of Toronto, 60 St.~George St., Toronto, Ontario, M5S 1A7, Canada}
\author{Yong Baek Kim}
\affiliation{Department of Physics and Center for Quantum Materials, University of Toronto, 60 St.~George St., Toronto, Ontario, M5S 1A7, Canada}
\affiliation{Canadian Institute for Advanced Research, Toronto, Ontario, M5G 1Z8, Canada}
\author{Hae-Young Kee}
\affiliation{Department of Physics and Center for Quantum Materials, University of Toronto, 60 St.~George St., Toronto, Ontario, M5S 1A7, Canada}
\affiliation{Canadian Institute for Advanced Research, Toronto, Ontario, M5G 1Z8, Canada}

\begin{abstract}
  \noindent
  The spin liquid phase is one of the prominent strongly interacting topological phases of matter whose unambiguous confirmation is yet to be reached despite intensive experimental efforts on numerous candidate materials.
  Recently, a new family of correlated honeycomb materials, in which strong spin-orbit coupling allows for various bond-dependent spin interactions, have been promising candidates to realize the Kitaev spin liquid.
  Here we study a model with bond-dependent spin interactions and show numerical evidence for the existence of an extended quantum spin liquid region, which is possibly connected to the Kitaev spin liquid state.
  These results are used to provide an explanation of the scattering continuum seen in neutron scattering on $\alpha$-RuCl$_3$.
\end{abstract}
\maketitle

\noindent
{\bf Introduction}

The role of strong interaction between electrons in the emergence of topological phases of matter is currently a topic of intensive research.
The archetypal example of a topological phase with strong electron interaction is the quantum spin liquid\cite{Balents2010}, in which the elementary excitations are charge-neutral fractionalized particles.
While a lot of progress has been made on the theoretical understanding of the quantum spin liquid phase, its direct experimental confirmation has remained elusive despite various studies on a number of candidate materials\cite{Shimizu2003,Helton2007,Okamoto2007,MYamashita2010,Han2012}. 
Significant progress, however, has recently been made due to the availabilty of a new class of correlated materials, where strong spin-orbit coupling leads to various bond-dependent spin interactions\cite{Jackeli2009,WCKB2013,RLK2016}, thus resulting in magnetic frustation.
These materials are Mott insulators with 4$d$ and 5$d$ transition metal elements, which include iridates and ruthenates\cite{Singh2012,Plumb2014,Modic2014,Takayama2015} and come in two- or three-dimensional honeycomb variants.
They have been particularly exciting because they intrinsically have a strong Kitaev interaction and therefore could potentially realize the Kitaev spin liquid (KSL) phase -- an example of a $\mathbb{Z}_2$ quantum spin liquid where the electron's spin-$\frac{1}{2}$ fractionalizes into two degrees of freedom: itinerant Majorana fermions and $\mathbb{Z}_2$ fluxes.

While the Kitaev interaction ($K$) in these materials is large, it competes with symmetry allowed nearest-neighbor (n.n.) symmetric off-diagonal ($\Gamma$) and Heisenberg ($J$) spin interactions\cite{Rau2014}.
For example, in $\alpha$-RuCl$_3$ (RuCl$_3$), an actively studied KSL candidate, comprehensive {\it ab initio} computations and recent dynamical studies\cite{Ran2017,Wang2017} suggest that ferromagnetic $K$ and antiferromagnetic $\Gamma$ interactions are dominant and comparable in magnitude, while $J$ is negligible\cite{Kim2016,Winter2016,Yadav2016}.
The balance of these and additional small further neighbor interactions causes RuCl$_3$ and other KSL candidates to magnetically order at low temperature; however, it is still unclear whether or not the often-large $\Gamma$ interaction prefers magnetic ordering.
Meanwhile, the community has attempted to revive the possibility of a KSL in RuCl$_3$ by applying a small magnetic field, with the effect of entering a potential spin liquid phase with no magnetization\cite{Yadav2016}.

\begin{figure}[!h]
  \centering
  \includegraphics[width=\linewidth]{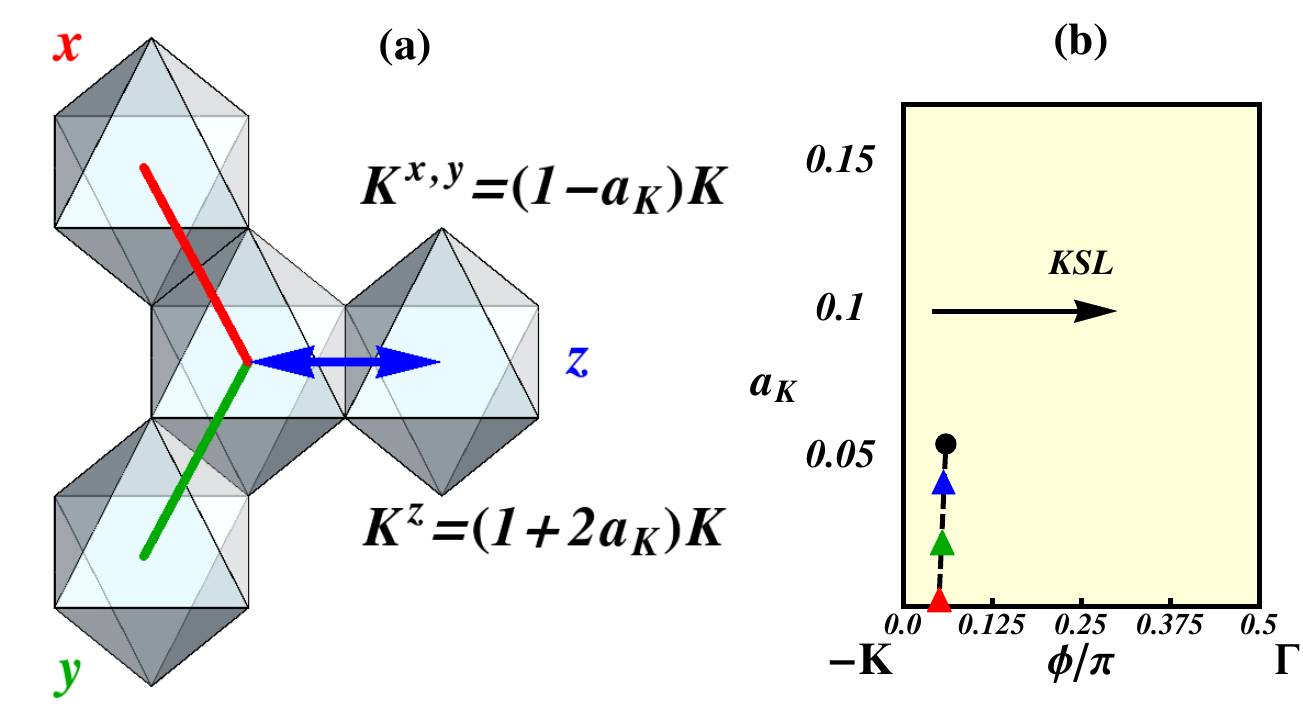}
  \caption{(Color online) (a) Shorter z-bond leading to stronger $K^z$ and weaker $K^{x,y}$ interactions, parameterized by $a_K$ in Eq. (\ref{HGK}). (b) $a_K$ phase diagram of Eq. (\ref{HGK}).
    A line of first order phase transitions (dashed black line), terminating below $a_K = 0.06$ (black dot), separates the $-K$ and $\Gamma$ limits.
    Anisotropy allows the KSL to be adiabatically connected to the $\Gamma$-phase (arrow).
    Red, green and blue triangles indicate first-order phase transitions when $a_K = 0.0, 0.02$ and $0.04$ respectively, as seen in $N = 18$ and 24 site exact diagonalization calculations.
    A subsequent iDMRG calculation\cite{Gohlke2017} has further corroborated the phase transition labeled by the red triangle.
  }
  \label{fig:anipd}
\end{figure}
Since a weak magnetic field takes RuCl$_3$ out of the ordered phase, it lends credence to the idea that the zig-zag phase is stabilized by small interactions at comparable energy scale to the magnetic field, such as a 3rd n.n. Heisenberg $J_3$\cite{Kim2016,Winter2016,Yadav2016} term or terms coming from slight trigonal distortion\cite{RauKeeTD}.
This calls into question the role of the $\Gamma$ interaction.
Interestingly, a recent analysis of the $\Gamma$ model revealed a macroscopically degenerate classical ground state\cite{rousochatzakis_classical_2016}.
    
In this work we will thus investigate if a model with $K$ and $\Gamma$ hosts an extended quantum spin liquid phase.
A previous exact diagonalization study on a 24 site honeycomb cluster hints that the ferromagnetic KSL is unstable after perturbing with a small $\Gamma$, but the resulting phase is not orderered\cite{Rau2014}.
On the other hand, it is known that the KSL is stable upon introducing bond anisotropy, which is present in real materials as depicted in Fig. \ref{fig:anipd}a.
We indeed find that such anisotropy can extend the KSL phase between the $-K$ and $\Gamma$ limits, as shown in Fig. \ref{fig:anipd}b.

\begin{figure*}
  \centering
  \includegraphics[width=\textwidth]{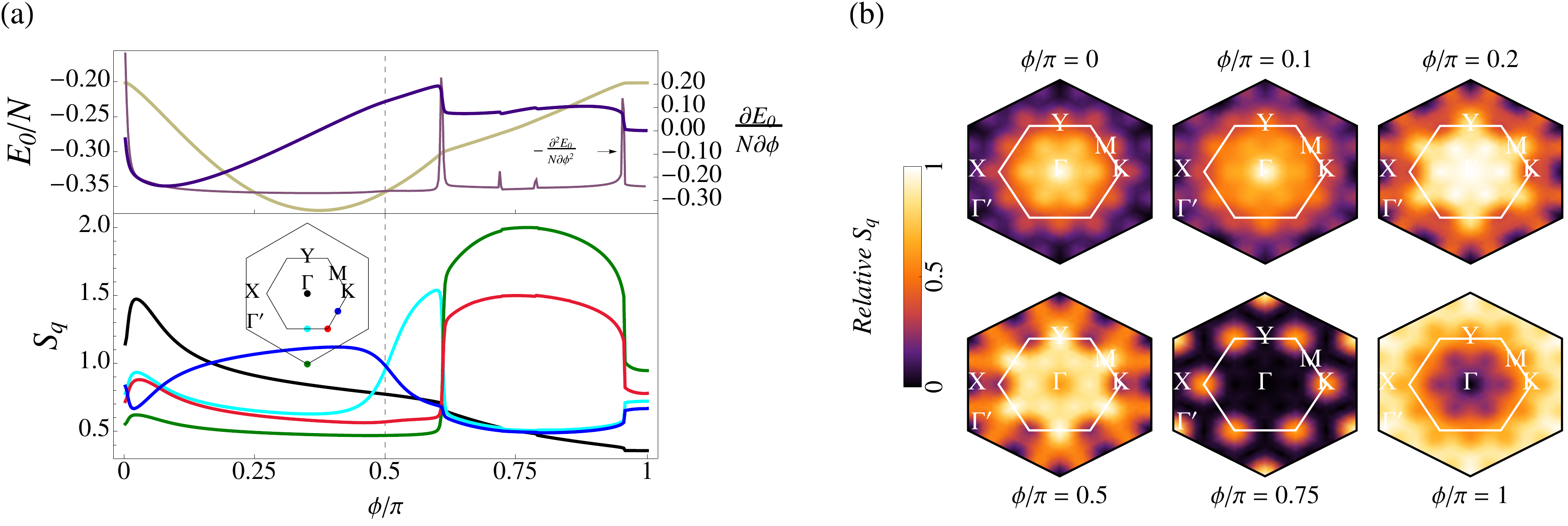}
  \caption{(Color online) (a) {\it Top:} $E_0/N$ (yellow), $\frac{1}{N}\frac{\partial E_0}{\partial\phi}$ (purple) and $-\frac{1}{N}\frac{\partial^2 E_0}{\partial\phi^2}$ (light purple) for anisotropy parameter $a_K=0.1$. {\it Bottom:} $S_q$ for anisotropy parameter $a_K=0.1$ at $\Gamma$- (black), M- (blue), Y- (cyan), K- (red) and $\Gamma^\prime$- (green) in the reciprocal lattice (inset). (b) Representation of $S_q$, averaged over domains in a real material, when $a_K=0.1$ for various $\phi$ in the phase diagram.}
  \label{fig:Fig1}
\end{figure*}
We consider the following nearest-neighbor (n.n.) model on a two-dimensional honeycomb lattice:
\begin{equation}
  H = \sum_{\gamma\in x,y,z} H^\gamma,\label{HGK}
\end{equation}
where
\begin{equation}
  H^z = \sum_{\langle ij\rangle\in z-bond} [K_z S^z_iS^z_j + \Gamma_z(S_i^xS_j^y+S^y_iS^x_j)]
\end{equation}
and $H^{x,y}$ are defined similarly with corresponding $K_{x,y}$ and $\Gamma_{x,y}$.
Each $H^\gamma$ represents the n.n spin interactions along one of the three bond directions, $\gamma = x, y, z$. The model is parameterized by $K_z = -(1 + 2a_K)\cos\phi$, $K_{x,y} = -(1-a_K)\cos\phi$, $\Gamma_{x,y,z} = \sin \phi$, with $a_K$ characterizing bond anisotropy. When $\phi = 0,\pi$ (i.e. $\Gamma_\gamma = 0$), this model reduces to the exactly solvable Kitaev model with the KSL ground state.



  We have studied this model using a combination of three different, corroborating, numerical methods: exact diagonalization (ED) on a 24-site honeycomb cluster, the method of thermal pure quantum states\cite{JPSJ.55.3354,PhysRevB.49.5065,PhysRevE.62.4365,SSugiura2012,SSugiura2013}, and infinite time-evolution block decimation (iTEBD).
  We have first reproduced the earlier work in the isotropic $a_K = 0$ limit, showing a strong first-order transition between $-K$ and $\Gamma$ limits (see Supplementary Materials (SM)).
  We present the following results using our numerical techniques:
\\
1) When $a_K \geq 0.06$, we find that the $-K_\gamma$ ($0 \leq \phi \leq \pi/2$) and $\Gamma_\gamma$ ($\phi/\pi = 0.5$) limits are adiabatically connected as shown in Fig. \ref{fig:anipd}b. Thus we find evidence for an extended quantum spin liquid phase in the presence of anisotropy, $a_K$.
\\
2) An intervening magnetically ordered phase separates the spin liquid phase near the pure $\Gamma_\gamma$ limit and the antiferromagnetic KSL at $\phi/\pi=1$.\\
3) The specific heat $C(T)$ and entropy $S(T)$ at finite temperatures
suggest a smooth crossover from the ferromagnetic Kitaev limit to the pure $\Gamma_\gamma$ limit, consistent with our ED results. \\
4) Zig-zag spin correlations become dominant upon perturbing the quantum spin liquid phase in $0 < \phi < \pi/2$ by $J_3$, indicating the enhancement of zig-zag order by $J_3$. 

\noindent
\\~\\
{\bf Results} \\
{\it Extended spin liquid state in global phase diagram ---}
The ground state energy per site $E_0/N$ of Eq. \eqref{HGK} was computed for $\phi/\pi \in [0,1]$, and for different anisotropy parameters  
by ED on a 24-site cluster using periodic boundary conditions (see SM).
Discontinuities in $\frac{1}{N}\frac{\partial E_0}{\partial\phi}$ were used to identify possible phase transitions.
Remarkably, when $\phi/\pi \in [0,0.5]$ and $0 \leq a_K < 0.06$, there is a line of first order phase transitions that terminate at $a_K = 0.06$. Above $a_K = 0.06$, the first derivative of the energy presents no sharp features suggesting that {\it the ground state of the $\Gamma$-limit} ($\phi/\pi=0.5$) {\it is adiabatically connected to the ferromagnetic Kitaev spin liquid} ($\phi/\pi=0$) as depicted in Fig. \ref{fig:anipd}b for $a_K = 0.1$.

In the antiferromagnetic region of phase space, there are two large discontinuities in $\frac{1}{N}\frac{\partial E_0}{\partial\phi}$ that encompass a large region of phase space separating the $\Gamma$-limit and the exactly solvable antiferromagnetic Kitaev limit at $\phi/\pi = 1$.
These peaks coincide with kinks in $E_0/N$ (solid yellow) shown in Fig. \ref{fig:Fig1}a.
Two smaller discontinuities can also be seen near $\phi/\pi = 0.75$, however these are not present when $a_K=0$, while the larger jumps near $\phi/\pi = 0.5$ and $1$ appear consistently for different $a_K$.
The small discontinuities can thus be considered spurious and a consequence of the finite cluster size. 
Similar finite size effects were also found for $\phi/\pi \in [0,0.5]$ when $a_K=0$, as discussed in the SM.

{\it Magnetic order and perturbations --- }The ground state wavefunction of Eq. \eqref{HGK} computed by ED is used to evaluate real-space spin-spin correlation functions $\langle{\bf S}_i\cdot{\bf S}_j\rangle$, where $i$ and $j$ are site indices on the honeycomb lattice.
By Fourier transform, we obtain the static structure factor (SSF) given by $S_{\textbf{q}} = \frac{1}{N}\sum_{i,j} e^{i(\textbf{r}_i-\textbf{r}_j)\cdot \textbf{q}}\langle{\bf S}_i\cdot{\bf S}_j\rangle$ where $\textbf{q}$ is a vector in the reciprocal lattice.
The SSF at various points in the Brillouin zone (BZ) is plotted over the phase space in the bottom panel of Fig. \ref{fig:Fig1}a. 

The discontinuities in the SSF can be directly matched with those in $\frac{1}{N}\frac{\partial E_0}{\partial\phi}$.
Visualizations of the SSF over the BZ for representative $\phi$ in the phase diagram are presented in Fig. \ref{fig:Fig1}b.
The SSF in Fig. \ref{fig:Fig1}b is obtained by computing the average of $\langle S_i\cdot S_j\rangle$ over all n.n. bonds, 2nd n.n., etc. 
This calculation reflects the presence of different domains in the crystal, in which either of $x,y,z$ bond interactions can be stronger and thus, over the whole crystal, these domains result in an isotropic $S_q$ despite the inherent bond anistropy in Eq. \eqref{HGK}.
The SSF varies adiabatically when $a_K=0.1$ for $\phi \in [0,\pi/2]$ and the spin correlations at the $\Gamma$- and M-points are comparable in intensity when $\Gamma \simeq K_\gamma$, leading to a ``star''-shaped structure in the SSF as seen in Fig. \ref{fig:Fig1}b (e.g, $\phi/\pi = 0.2$). 
The extended phase separating $\phi/\pi=0.5$ and $1$ is characterized by dominating spin correlations at the K- and $\Gamma^\prime$-points in the reciprocal lattice ($\phi/\pi = 0.75$ in Fig. \ref{fig:Fig1}b).
Contained within this phase is the exactly solvable point with hidden $SU(2)$ symmetry at $\phi/\pi=0.75$ which features K-point correlations\cite{Chaloupka2015} consistent with the results presented here.
Thus the extended spin liquid phase for ferromagnetic K is separated from the antiferromagnetic KSL at $\phi/\pi = 1$ by a magnetically ordered phase.

\begin{figure}[!h]
  \centering
  \includegraphics[width=\linewidth]{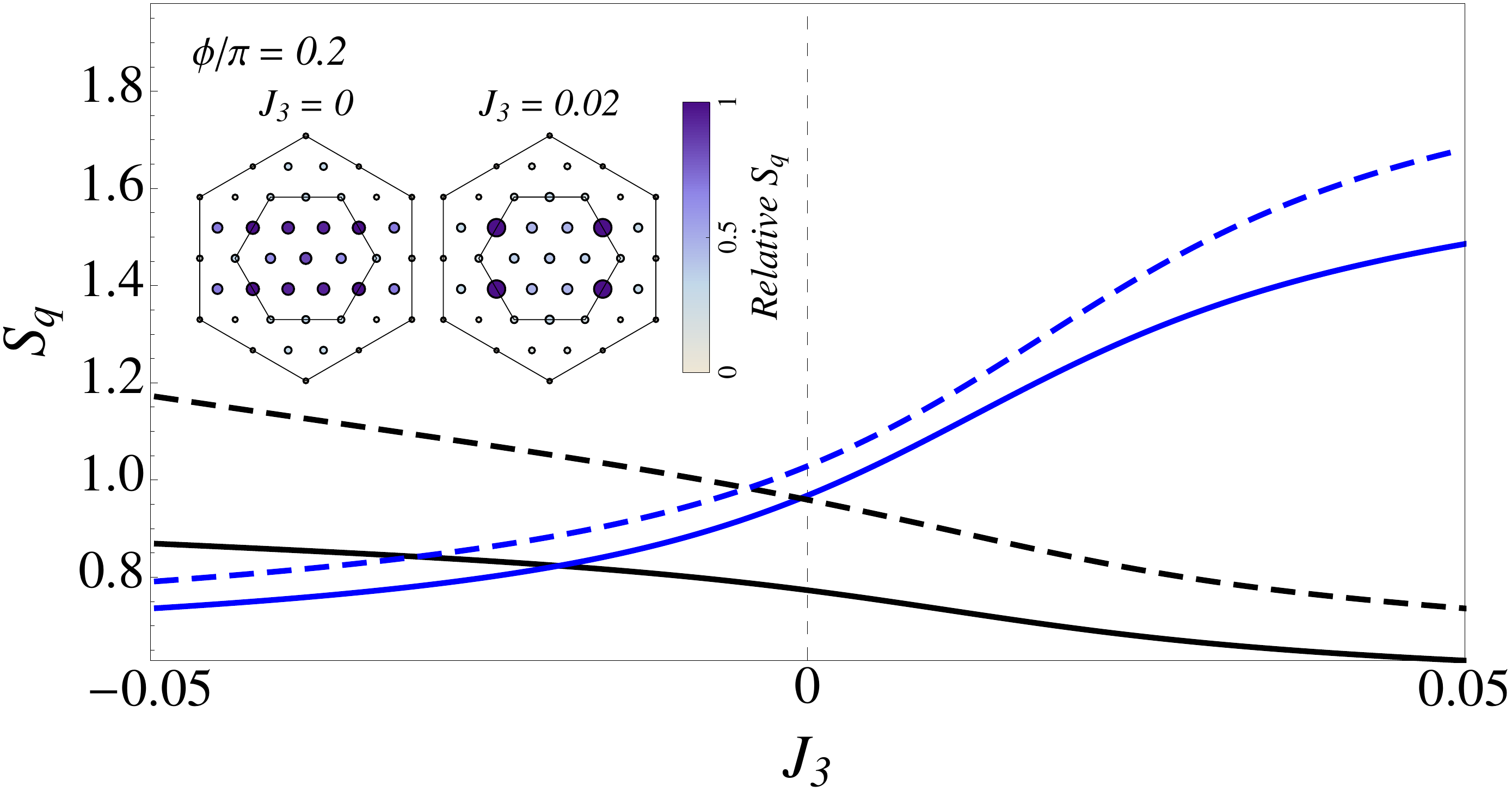}
  \caption{(Color online) SSF at the $\Gamma$- (black) and $M$-points (blue) in the BZ for small $J_3$. Solid and dashed curves correspond to $\phi/\pi=0.5$ and $0.2$ respectively. The inset shows the dramatic change in the relative intensity at the $\Gamma$- and $M$-points for small $J_3 > 0$.}
  \label{fig:Fig3}
\end{figure}

These results can be connected to real materials, particularly RuCl$_3$ in which a zig-zag magnetic ordering has been observed\cite{Jennifer2015,Johnson2015,Cao2016}.
Previous studies have shown that in addition to the n.n. ferromagnetic Kitaev and antiferromagnetic $\Gamma$ interactions, a 3rd n.n. antiferromagnetic Heisenberg interaction $J_3\sum_{\langle\langle\langle i,j\rangle\rangle\rangle} {\bf S}_i \cdot{\bf S}_j$ is non-vanishing and plays a role in determining the magnetic ordering in RuCl$_3$\cite{Kim2016,Winter2016}.
Fig. \ref{fig:Fig3} shows that the effect of perturbing Eq. \eqref{HGK} by $J_3$ is to enhance (suppress) the M-point ($\Gamma$-point) spin correlations, consistent with a zig-zag magnetically ordered state observed in experiments. 
This result indicates that by tuning $J_3$ in the real material, an {\it alternate path to achieve a spin liquid phase may be realized}.

{\it Specific heat and thermal entropy ---} Previous study on the finite temperature properties of the Kitaev model has shown that Kitaev spin liquids feature two peaks in the heat capacity $C(T)$ and a $\frac{1}{2}$-plateau in the entropy $S(T)$, which is attributed to the thermal fractionalization of spin degrees of freedom\cite{JNasu2015}.
Here we go beyond the Kitaev limit and investigate the heat capacity and entropy at finite temperature in the presence of $\Gamma$, which is expected to compete with $K_\gamma$ in RuCl$_3$, using the method of thermal pure quantum states (see SM). 

\begin{figure}[!h]
  \includegraphics[width=\linewidth]{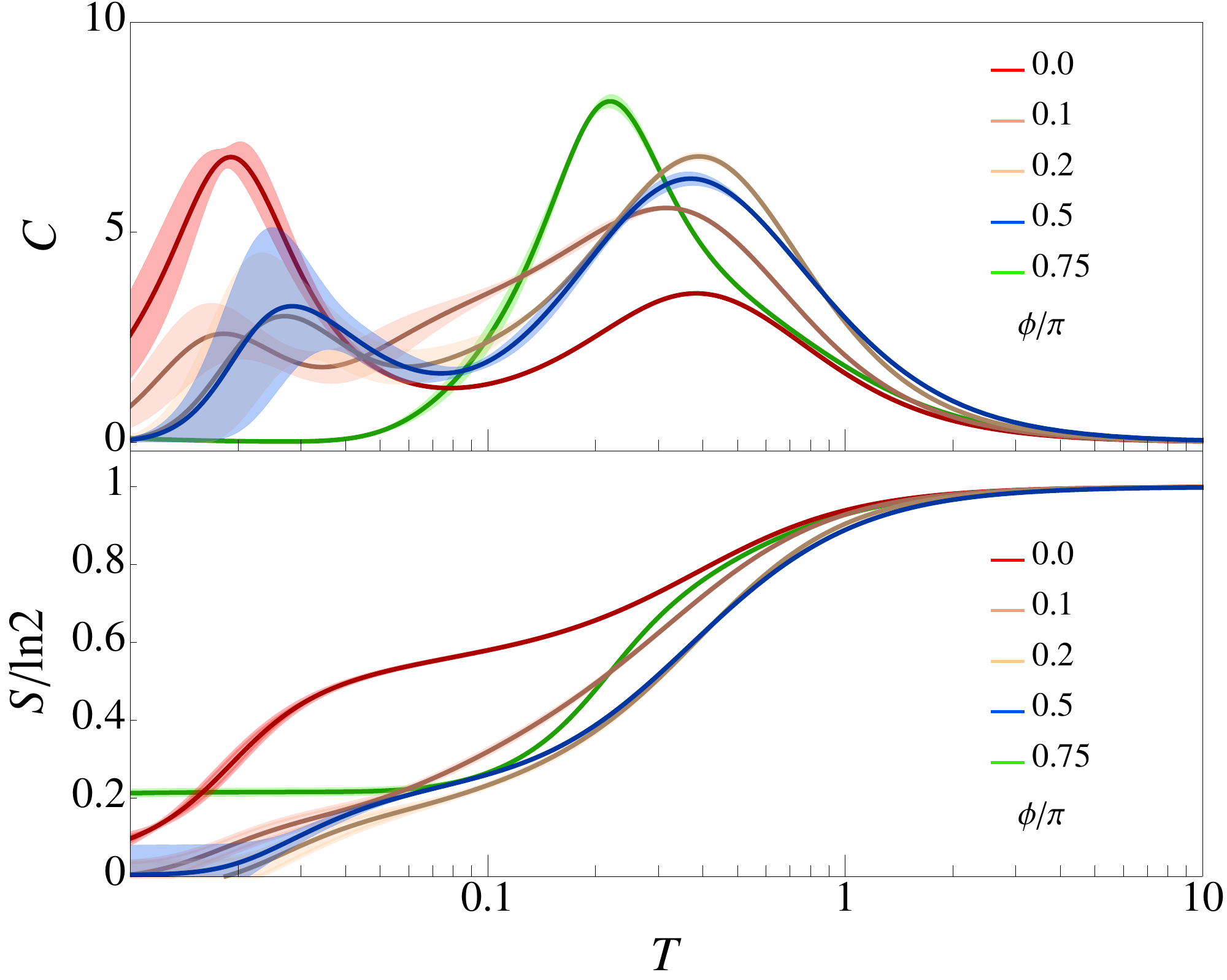}
  \caption{(Color online) Results of the method of thermal pure quantum states on a 24-site cluster with anisotropy parameter $a_K=0.1$. Plotted in solid curves are the temperature dependence of heat capacity $C(T)$ (top) and entropy $S(T)$ (bottom) for various $\phi$ in the phase diagram. The shaded regions represent the estimated errors on the results. The temperature $T$ is expressed in units where $\sqrt{(K_z/(1+2a_K))^2+\Gamma_z^2}=\sqrt{(K_{x,y}/(1-a_K))^2 + \Gamma_{x,y}^2}=1$.
  }
  \label{fig:Fig4}
\end{figure}

The dependence of $C(T)$ on $\phi$ when $a_K=0.1$ is plotted in the top panel of Fig. \ref{fig:Fig4}. The expected two peak structure in $C(T)$ is observed when $\phi/\pi = 0$, and is seen to be maintained continuously as $\phi/\pi$ approaches $0.5$ so that the $\Gamma$-limit shows a qualitatively similar behaviour in $C(T)$ to the Kitaev spin liquid. 
Evidence for a phase transition can be seen when $\phi \gtrsim 0.7$ on account of the abrupt change in $C(T)$, resembling that of the heat capacity in trivially ordered phases \cite{PhysRevB.93.174425}. This finding is consistent with our ED results and the 120-order at $\phi/\pi=0.75$ seen in Refs. \onlinecite{Rau2014} and \onlinecite{Chaloupka2015}.
The dependence of $S(T)$ on $\phi$ is plotted in the bottom panel of Fig. \ref{fig:Fig4} with a clear $\frac{1}{2}$-plateau observed when $\phi/\pi = 0$, consistent with the expected Kitaev spin liquid behaviour. In addition, a plateau of about 1/5 the total entropy is observed when $\phi/\pi = 0.5$. Another plateau is observed in the magnetically ordered phase around $\phi/\pi = 0.75$; however, this feature can be attributed to finite-size effects as follows. The $(N+1)$-fold ground state degeneracy at $\phi/\pi = 0.75$ due to the hidden $SU(2)$ symmetry\cite{Chaloupka2015} is only slightly broken away from this point, inducing a plateau in $S(T)$ with height given by $\ln (N+1)/N \ln 2 \simeq 0.1935 \sim 1/5$ when $N=24$. By contrast, the height of the plateau around $\phi/\pi = 0.5$ is independent of $N$\cite{Note1}.

The physical origin of the two peak structure in $C(T)$ and the plateau in $S(T)$ can be traced to the energy scales of the thermal fluctuations of the underlying quasiparticles in the spin liquid \cite{JNasu2015}.
In the Kitaev spin liquid at zero temperature, the low-lying quasiparticle excitations are characterized by itinerant Majorana fermions which disperse in a background of zero flux\cite{Kitaev2006,Knolle2014}.
It has been shown that as temperature increases, the flux degrees of freedom begin to fluctuate and lead to the low temperature peak seen in $C(T)$, resulting in the plateau seen in $S(T)$. 
Furthermore, the high temperature peak in $C(T)$ is attributed to the development of short range spin correlations\cite{JNasu2015}. 
Our results show that the two-peak structure in $C(T)$ 
is qualitatively maintained and further suggests that no phase transition has taken place.

{\it Similarities on the infinite tree ---}
We further studied Eq. \eqref{HGK} on an infinite Cayley tree with $z=3$ connectivity, using the infinite time-evolving block decimation algorithm\cite{vidal_classical_2007} (iTEBD; see SM).
Classically, the ground state in the $\Gamma$-limit on the infinite tree is macroscopically degenerate because
a different state with the same energy can be constructed by flipping the sign of one spin component on an infinite string of neighboring spins.
The $\Gamma$-limit on the two-dimensional (2D) honeycomb and three-dimensional (3D) hyper-honeycomb\cite{rousochatzakis_classical_2016} lattices also feature similar classical degeneracy.
The similarity at the classical level of the $\Gamma$-limit on the infinite tree to the 2D and 3D lattices prompts us to study the quantum model on the infinite tree for further insight.

\begin{figure}
  \includegraphics[width=\linewidth]{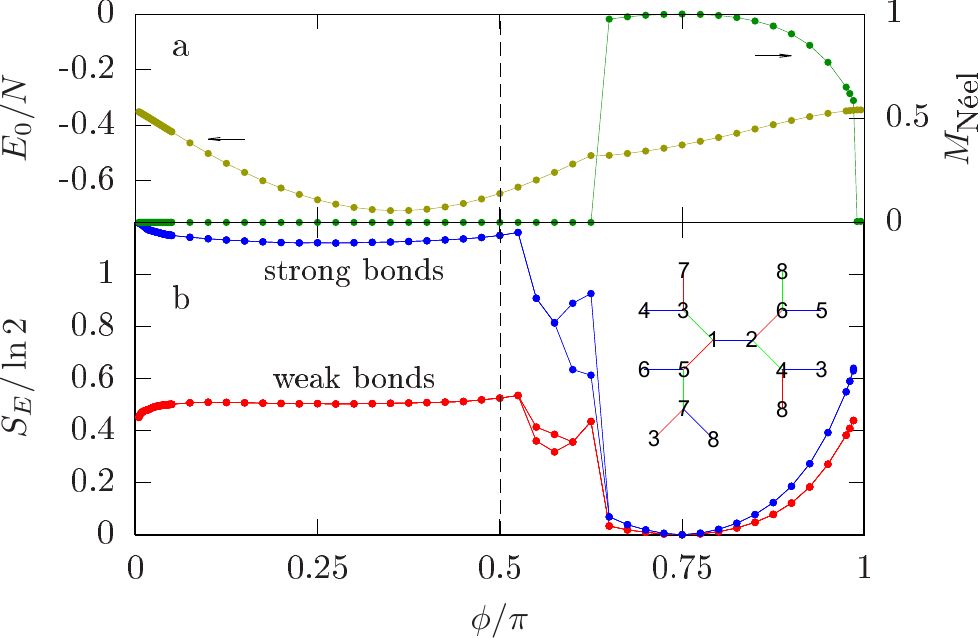}
  \caption{(Color online) iTEBD results for small anisotropy $a_K=a_\Gamma=0.1$, and bond
    dimension $\chi=10$. (a) N\'eel order parameter,
    $M_{\mbox{N\'eel}}$ and ground state energy per site, $E_0/N$. 
    (b) Entanglement entropy $S_E$ associated with splitting the system along the different bonds in
    the eight-site unit cell. In most of the phase diagram all the
    strong $z$ bonds are identical, and so are all the weak, $x$ and
    $y$ bonds. In the transition region near $\phi=0.6\pi$, the
    symmetry between like bonds is broken, perhaps indicating a first
    order transition. At $\phi/\pi=0.75$, the entanglement
    entropy vanishes since the system is in a product state.
    Inset: schematic of Cayley tree with $z=3$ connectivity.}
  \label{fig:Fig5}
\end{figure}

Figure \ref{fig:Fig5} shows results of the eight-site iTEBD calculation with bond dimension $\chi=10$, and anisotropy $a_K=0.1$.
In this calculation, we have also introduced an anisotropy to $\Gamma_\gamma$ such that $\Gamma_x = \Gamma_y = (1-a_\Gamma)\sin\phi$ and $\Gamma_z = (1+2a_\Gamma)\sin\phi$ in order to apply the iTEBD method (see SM).
No transition is found when $\phi/\pi \in [0,0.5]$ and
the obtained state is a highly entangled paramagnet, with $S_E\sim 0.8$ for strong ($z$) bonds, while for weak ($x,y$) bonds, $S_E \sim 0.4$.
Deep in the gapped phase of the Kitaev model, with large anisotropy $a_K$, one finds $S_E\sim\log 2\sim 0.693$ for the strong bonds and much smaller values of $S_E$ for the weak bonds.
Both, however, increase as the anisotropy is reduced - perhaps due to a finite contributions from the Majorana fermions\cite{kimchi_three-dimensional_2014}.
An increase is expected in the entanglement entropy as one approaches a phase transition, 
however no such peaks are seen for $0<\phi<\pi/2$.
Similarly, there are no sharp features in the ground state energy $E_0$ as a function of $\phi$, which indicates that this phase is adiabatically connected to the Kitaev spin liquid at $\phi=0$.

There is an apparent first order transition around $\phi=0.6\pi$ into a N\'eel state with spins ordered in the [111] direction, accompanied by a dramatic lowering of $S_E$ on both strong and weak bonds into this region. 
This N\'eel state becomes a simple product state when $\phi/\pi=0.75$, as seen by the vanishing of $S_E$.
A final transition into a paramagnetic state is seen before the antiferromagnetic Kitaev limit. 

\noindent
\\~\\
{\bf Discussion }
\\~\\
The highlight of our numerical results is that, in the presence of bond anisotropy $a_K$, there exists an extended quantum spin liquid region which is adiabatically connected to the ferromagnetic KSL.
The model we have studied is motivated by experiments on RuCl$_3$ and earlier {\it ab initio} computations\cite{Kim2016,Winter2016,Yadav2016}.
In a recent inelastic neutron scattering experiment on RuCl$_3$, it is found that the continuum of finite energy excitations exists both below and above the magnetic transition temperature despite that the low temperature ground state is the zig-zag long-range ordered state\cite{Banerjee2016}.
The inelastic neutron scattering data for the continuum show the “star”-shape intensity that extends from the zone center towards the M points of the Brillouin zone.

Recall that the static structure factor in our ED study shows enhanced (decreased) short-range spin correlations at the M point (zone center) of the Brillouin zone as one moves from the ferromagnetic Kitaev limit to the pure $\Gamma_\gamma$ limit. When the strength of the ferromagnetic Kitaev interaction and the $\Gamma_\gamma$ interaction become comparable, both of the short-range spin correlations at the M and the zone center would show significant intensity, which leads to the “star”-shape structure in momentum space.
This behavior may be favorably compared to the finite-energy short-range spin correlations seen in RuCl$_3$.
Given that the {\it ab initio} computations suggest comparable magnitudes of the ferromagnetic Kitaev and $\Gamma_\gamma$ interactions in RuCl$_3$\cite{Kim2016}, it is conceivable that RuCl$_3$ may be very close to the quantum spin liquid phase found in our model and, as shown in our work, the introduction of small $J_3$ would favor the zig-zag magnetically ordered phase observed in RuCl$_3$.

Finally, more analytical understanding of the connection between the pure Kitaev limit and the quantum spin liquid phases identified in our numerical work would be extremely valuable for future applications on real materials.
Note that a possible incommensurate magnetic ordering cannot be ruled out due to finite cluster size.
  However, based on the results of a recent iDMRG study\cite{Gohlke2017}, no evidence of incommensurate ordering allowed by their momentum cuts was found.
  It is also interesting to note that quantum fluctuations do not lift the infinite ground state degeneracy of the classical model for positive $\Gamma$, while they may lead to incommensurate ordering for negative $\Gamma$\cite{rousochatzakis_classical_2016}.
Thus it is likely that the positive $\Gamma$ regime studied here possesses a spin liquid ground state, and the precise nature of the spin liquid is an excellent topic for future study.

\noindent
\\~\\
  {\bf Methods }
  \\~\\
  Our results were obtained using the combination of the three independent numerical techniques listed below.
\\~\\
  {\it Exact diagonalization ---}
ED was performed on a 24 site cluster with periodic boundary conditions.
This cluster allows all the symmetries present in the infinite honeycomb lattice and has been used reliably in previous related classical and quantum studies \cite{Chaloupka2010,Rau2014}.
The Hamiltonian given by Eq. (1) in the main text does not have the $U(1)$ symmetry associated with $S^z$ conservation, making it impossible to block diagonalize into magnetization sectors.
Therefore, the translational symmetry of the 24 site cluster was used to block diagonalize into different momentum sectors to gain more information about its energy spectrum.
The lowest energies and corresponding wavefunctions of each block were then numerically obtained using the Lanczos method. Further details and calculations can be found in the SM.
\\~\\
  {\it Thermal pure quantum states ---} We used the method of thermal pure quantum states\cite{SSugiura2012,SSugiura2013} in our specific heat and thermal entropy calculations. Details about the construction of thermal pure quantum states and the subsequent calculation of specific heat and entropy can be found in the SM.
  \\~\\
  {\it Infinite time-evolving block decimation algorithm ---} The Hamiltonian given by Eq. (\ref{HGK}) was studied on an infinite Cayley tree with $z = 3$ connectivity using the infinite time-evolving block decimation algorithm\cite{vidal_classical_2007}. Details about the method and the construction of the ground state can be found in the SM.

  \noindent
  \\~\\
{\bf Acknowledgements}
\\~\\
This work was supported by the NSERC of Canada and the Center for Quantum Materials at the University of Toronto. Y. Y. was supported by JSPS KAKENHI (Grant Nos. 15K17702 and 16H06345) and was supported by PRESTO, JST. Y. Y. was also supported in part by MEXT as a social and scientific priority issue (Creation of new functional devices and high-performance materials to support next-generation industries) to be tackled by using post-K computer. Computations were mainly performed on the GPC supercomputer at the SciNet HPC Consortium.  SciNet is funded by: the Canada Foundation for Innovation under the auspices of Compute Canada; the Government of Ontario;  Ontario Research Fund - Research Excellence; and the University of Toronto. A part of the TPQ results were checked by a program package, H$\Phi$\cite{Kawamura2017180}. We thank helpful discussions with Frank Pollmann, Matthias Gohlke, Shunsuke Furukawa, and Subhro Bhattacharjee. We particularly thank Natalia Perkins and Ioannis Rousochatzakis for informing us of their unpublished ED results on related models.
\noindent
\\~\\
  {\bf Contributions}
\\~\\
A.C. and Y.Y. performed the exact diagonalization calculations. Y.Y. performed the thermal pure quantum states calculations. G.W. performed the iTEBD calculations. H.-Y.K. and Y.B.K. supervised the study. All authors contributed to the writing of the manuscript.
\noindent
\\~\\
  {\bf Competing Interests}
  \\~\\
  The authors declare no competing financial or non-financial interests. 
\noindent
\\~\\
  {\bf Data Availability}
  \\~\\
  All relevant data is available from the corresponding author.

%

\pagebreak

\section*{Supplementary Materials}
\section{Exact Diagonalization}

\subsection*{Ground state energy and $S_q$ in the $a_K=0$ limit}

\begin{figure}[!h]
  \centering
  \includegraphics[width=\linewidth]{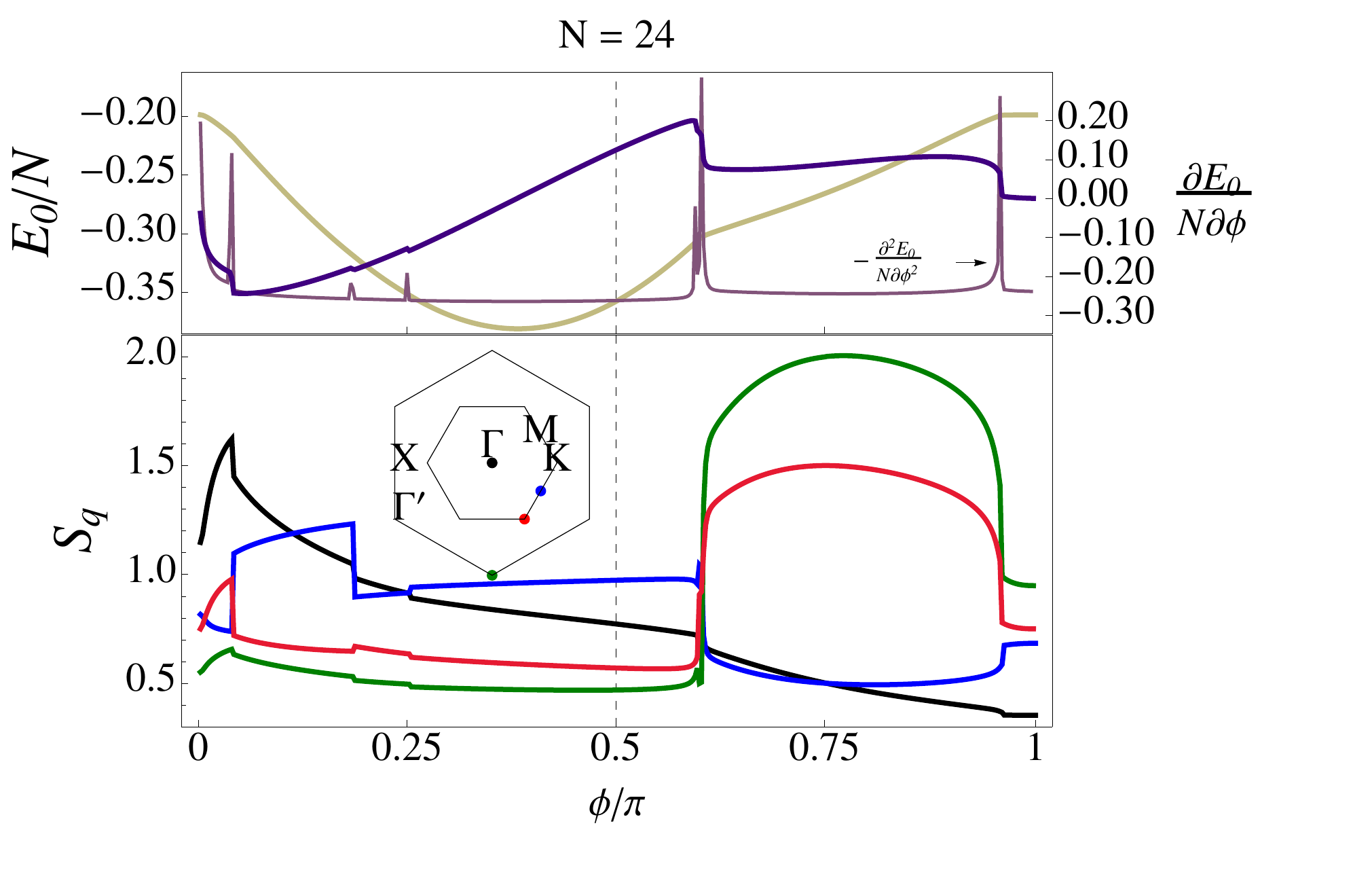}
  \caption{(Color online) {\it Top:} $E_0/N$ (yellow), $\frac{1}{N}\frac{\partial E_0}{\partial\phi}$ (purple) and $-\frac{1}{N}\frac{\partial^2 E_0}{\partial\phi^2}$ (light purple) for anisotropy parameter $a_K=0$ ($N = 24$). {\it Bottom:} $S_q$ for anisotropy parameter $a_K=0$ at $\Gamma$- (black), M- (blue), K- (red) and $\Gamma^\prime$- (green) in the reciprocal lattice (inset).}
  \label{fig:Fig1sm}
\end{figure}

When the Kitaev exchange is isotropic $a_K=0$, three discontinuities in $\frac{1}{N}\frac{\partial E_0}{\partial\phi}$, which are absent with slight anisotropy $a_K=0.1$, can be seen when $\phi/\pi \in [0,0.5]$.
These discontinuities are reflected in $S_q$ as can be seen in Fig. \ref{fig:Fig1sm}.
The general qualitative picture, however, is the same as with slight anistropy: comparable $\Gamma$- and M-point spin correlations when $\phi/\pi \in [0,0.5]$ lead to a ``star''-shaped pattern in $S_q$ in the Brillouin zone (BZ).
As when $a_K=0.1$, two large discontinuities in $\frac{1}{N}\frac{\partial E_0}{\partial\phi}$ and a dramatic change in $S_q$ when $K^\gamma > 0$ are evidence of an extended magnetically ordered phase with dominating K- and $\Gamma^\prime$-point correlations.
The inconsistent appearance of the other discontinuities in $\frac{1}{N}\frac{\partial E_0}{\partial\phi}$ (and associated small discontinuities in $S_q$) with changing anisotropy parameter are likely finite-size effect manifestations. Indeed, the location of these small discontinuities depends on system size ($N = 18, 24$). 

\begin{figure}[!h]
  \centering
  \includegraphics[width=\linewidth]{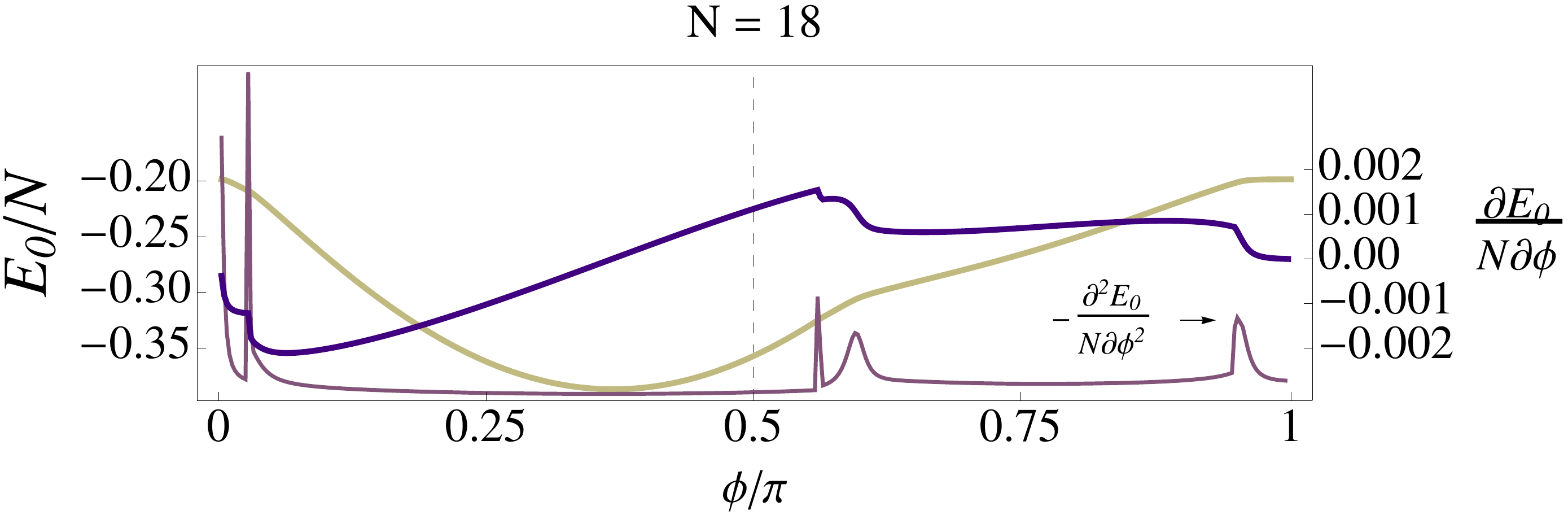}
  \caption{(Color online) $E_0/N$ (yellow), $\frac{1}{N}\frac{\partial E_0}{\partial\phi}$ (purple) and $-\frac{1}{N}\frac{\partial^2 E_0}{\partial\phi^2}$ (light purple) for anisotropy parameter $a_K=0$ ($N = 18$).
  }
  \label{fig:n18pd}
\end{figure}

\subsection*{Phase diagram for various $a_K \neq 0$}

\begin{figure}[!h]
  \centering
  \includegraphics[width=\linewidth]{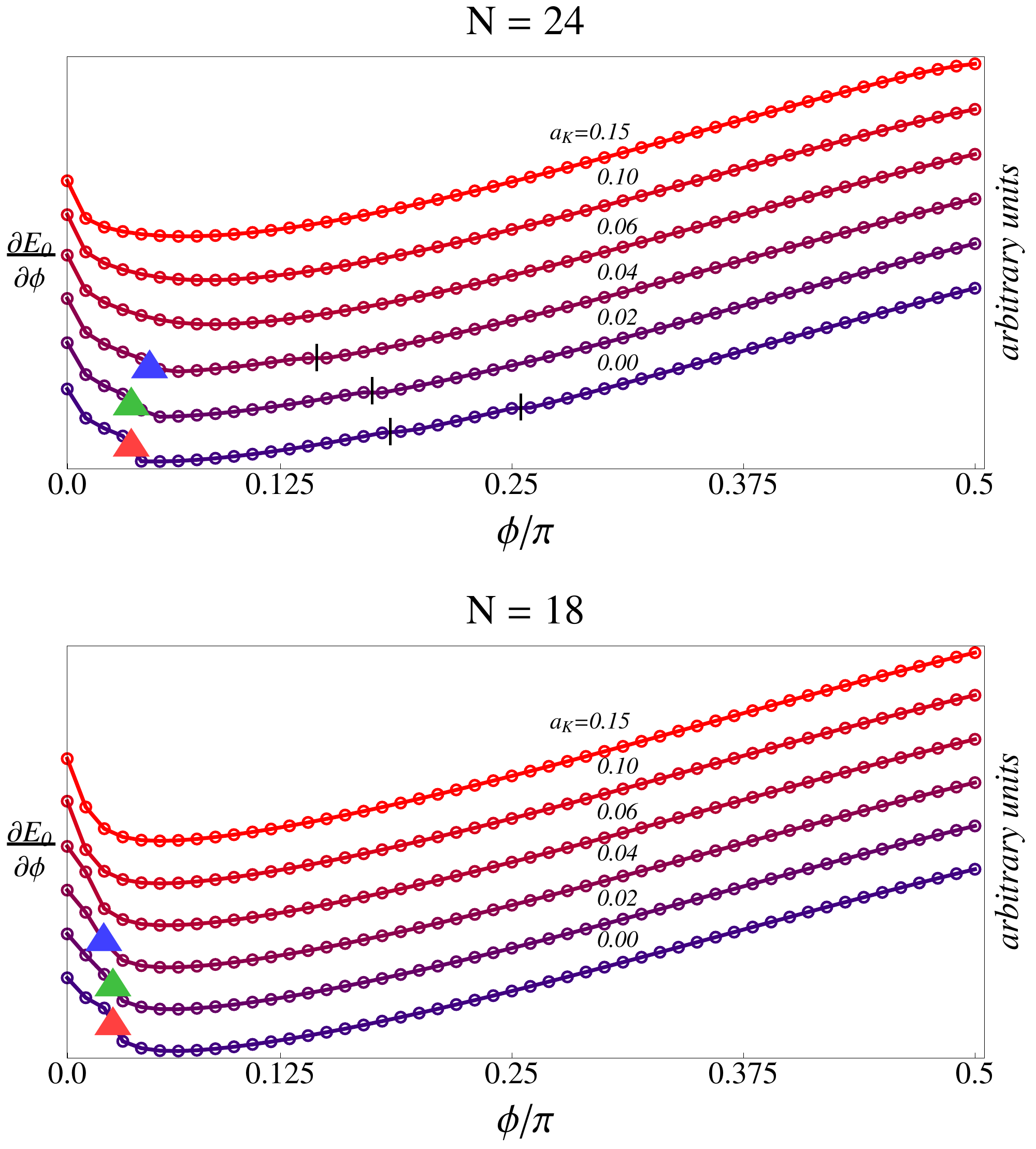}
  \caption{(Color online) $\frac{\partial E_0}{\partial \phi}$ for different anisotropy parameters when $\phi/\pi \in [0,0.5]$ ($N = 18, 24$).
    Red, green and blue triangles correspond to the locations of large first-order phase transitions (seen in both $N = 18$ and $24$ site clusters) depicted in Fig. 1b of the main text.
    Recent iDMRG calculations also corroborate the existence of the phase transition labeled by the red triangle\cite{Gohlke22017}.
    Black vertical bars label the location of small spurious discontinuities in $\frac{\partial E_0}{\partial \phi}$ which appear specifically when $N = 24$.
    These transitions do not appear when $N = 18$ nor in recent iDMRG calculations\cite{Gohlke22017}.
    We conclude that these transitions are likely due to finite-size effects, however, they do not affect our main conclusion, which is that the KSL and $\Gamma$ limits are adiabatically connected for small $a_K$.}
  \label{fig:n24apd}
\end{figure}

We investigated the phase diagram between $\phi/\pi = 0$ and $\phi/\pi = 0.5$ for different $a_K$ as shown in Fig. \ref{fig:n24apd} calculated by ED. The discontinuities seen in $\frac{\partial E_0}{\partial\phi}$ when $a_K = 0$ disappear for $a_K \geq 0.06$. The smooth connection between $\phi/\pi = 0$ and $\phi/\pi = 0.5$ limits is found to be robust for a wide range of $a_K$. 

\subsection*{Finite size spectrum}

The Hamiltonian was block diagonalized in terms of eigenvalues of the translation operators $T_{1,2}$ acting on the spin configuration on the 24 site cluster. In this notation, any spin configuration in the $S_z$ basis, $|s\rangle \equiv |S_1^z\dots S_{24}^z\rangle$, on the 24 site cluster can be translated as $T_1^nT_2^m|s\rangle$ where $n = 0, ... 5$ and $m = 0,1$. Periodic boundary conditions are imposed such that $T_1^6 = 1$ and $T_2^2 = 1$. Thus there are 12 unique eigenvalues of $T_1^nT_2^m$ (there are two atoms in the unit cell) of the form $e^{2\pi i (k_x/N_1 + k_y/N_2)}$, with $N_1 = 6$ and $N_2 = 2$ on the 24-site cluster. The momentum eigenvalues are labeled by the integer values $k_x$ and $k_y$. 

\begin{figure}[h]
  \includegraphics[width=\linewidth]{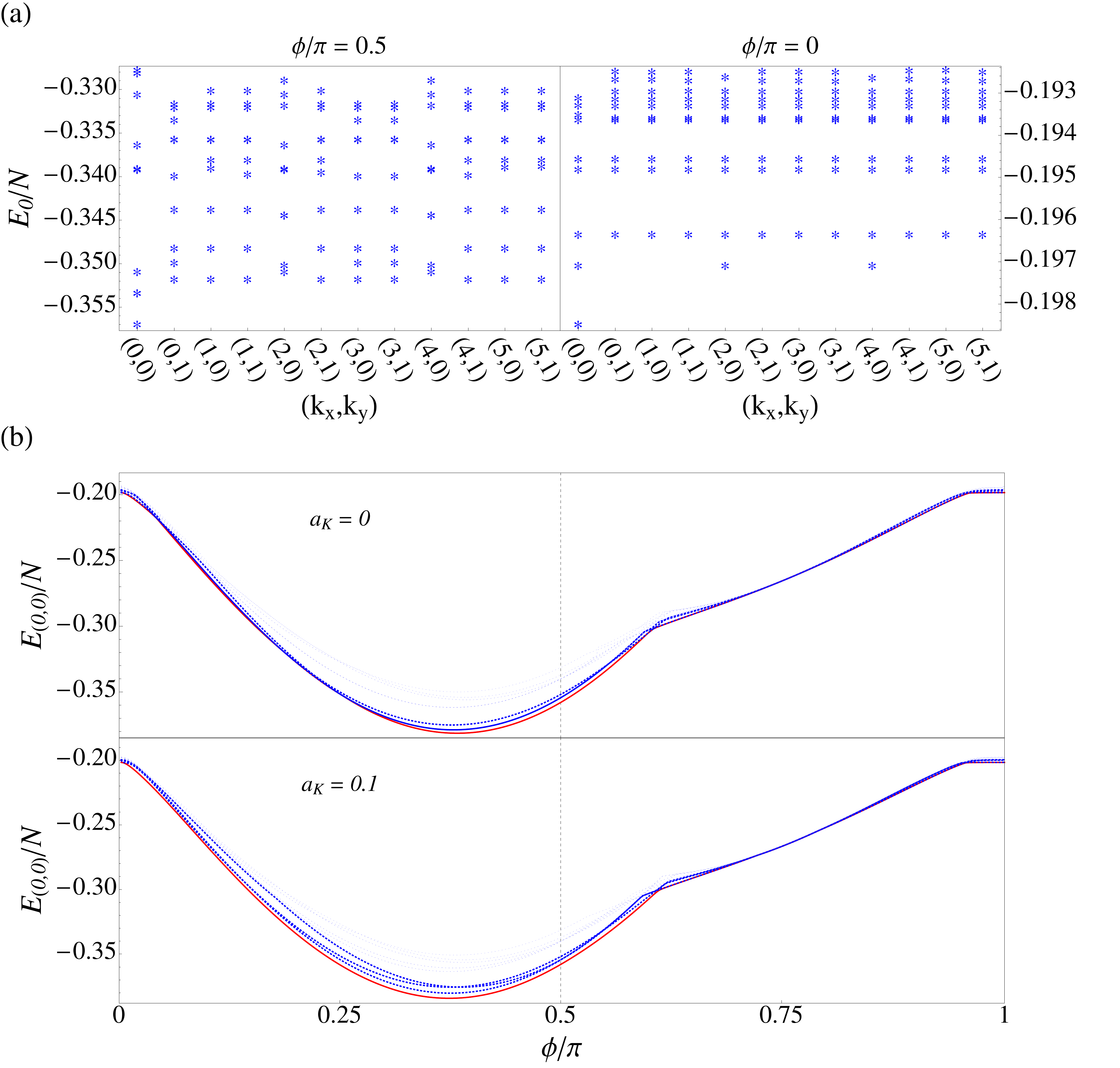}
  \caption{(Color online) (a) Energy spectrum across different momentum sectors in the $\Gamma$-limit (left panel) and $K$-limit (right panel) when $a_K = 0$. (b) Lowest eigenvalues in the $q = (0,0)$ momentum sector in the isotropic $a_K=0$ (top panel) and anisotropic $a_K=0.1$ (bottom panel). The ground state energy is highlighted in red. }
  \label{fig:Fig2sm}
\end{figure}

For all $\phi$, the ground state is in the $q \equiv (k_x,k_y) = 0$ momentum sector.
The energy spectra corresponding to $\phi/\pi = 0.5$ and $0$ are shown in Fig. \ref{fig:Fig2sm}a.
In the $\Gamma$-only limit, the ground state is accompanied close in energy by 3 excited states in the $q = 0$ sector, the first excited state being doubly degenerate.
These states are separated by a gap from the rest of the spectrum.
The Kitaev limit hosts a doubly degenerate ground state within ED.

It is found that when varying $\phi$ and keeping the Kitaev term isotropic, the four states in the $q=0$ sector experience level crossing, shown in Fig. \ref{fig:Fig2sm}b, and account for the first-order phase transitions found when $\phi \in [0,\pi/2]$ seen in Fig. \ref{fig:Fig1sm}.
The location of these level crossings when $\phi \in [0,\pi/2]$ depends on system size ($N = 18, 24$) and are thus attributed to finite size effect. 
By contrast, there is no level crossing when introducing small anisotropy $a_K=0.1$ to the Kitaev exchange.
The energy spectrum surrounding the exactly solvable point $\phi/\pi = 0.75$ tends to be degenerate and is qualitatively different from the spectrum in the rest of the phase space.
This result serves as a way to contrast the energy spectrum of the magnetically ordered phase with the rest of the phase diagram. 

\subsection*{$J_3$ perturbation}

\begin{figure}[h]
  \includegraphics[width=\linewidth]{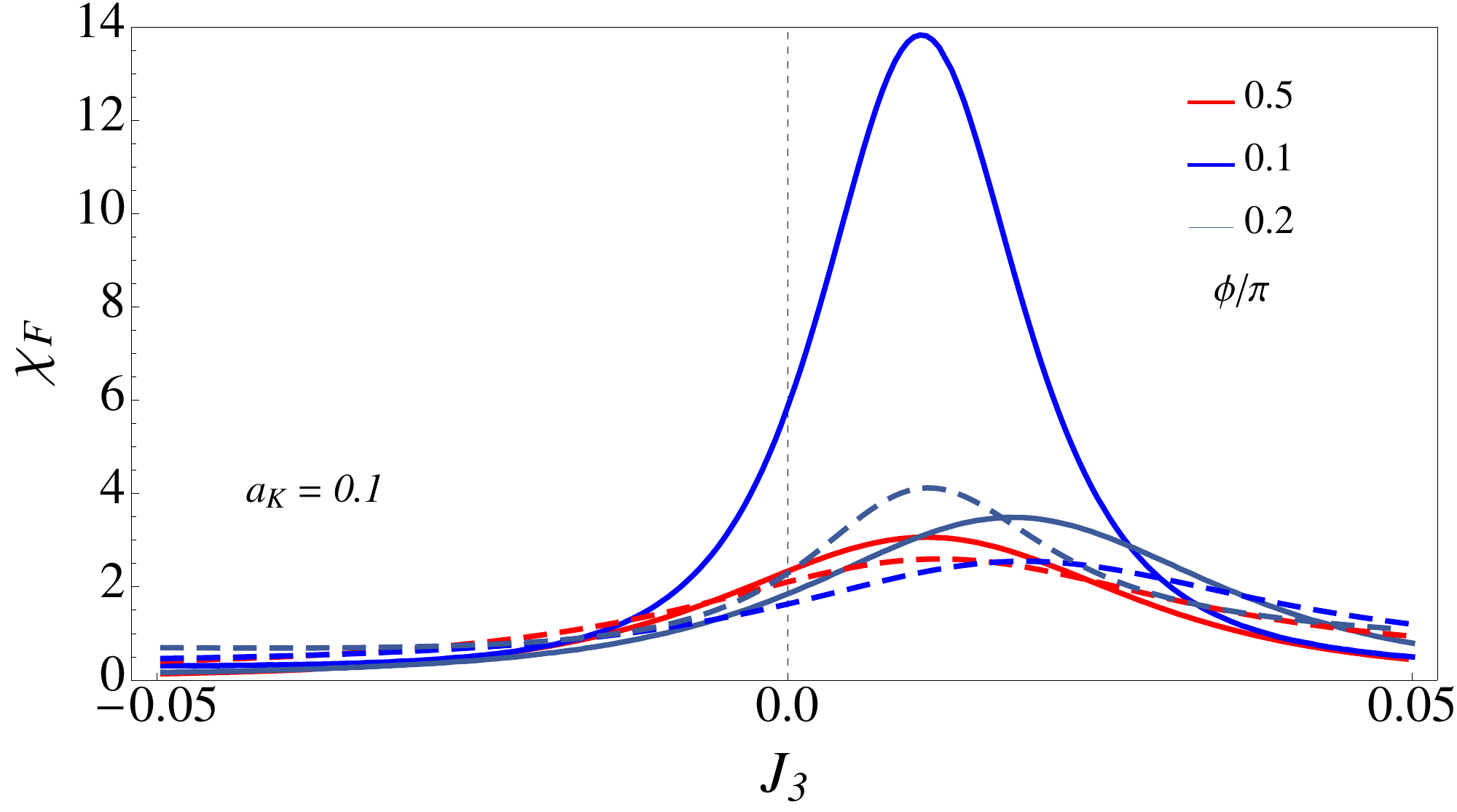}
  \caption{(Color online) Fidelity susceptibility (solid) and $-\frac{1}{N}\frac{\partial^2E_0}{\partial\phi^2}$ (dashed) for $\phi/\pi = 0.5, 0.1,$ and $0.2$ (red, blue, light blue) when $a_K=0.1$.}
  \label{fig:Fig3sm}
\end{figure}

Under the effect of a 3rd n.n. $J_3 > 0$, the intensity in $S_q$ at the M-points in the BZ was shown in the main text to be enhanced.
This result suggests that a phase transition into the experimentally observed zig-zag magnetically ordered state could occur for sufficiently large $J_3 > 0$.
Fig. \ref{fig:Fig3sm} shows signatures of a possible phase transition when investigating the second derivative of the energy (dashed lines) and fidelity susceptibility\cite{Albuquerque2010} (solid lines) $\chi_F$ when $\phi/\pi = 0.5, 0.1$ and $0.2$. 
The broad peaks in both $\chi_F$ and $-\frac{1}{N}\frac{\partial^2E_0}{\partial\phi^2}$ could be a result of finite size effect, and makes it difficult to estimate the necessary $J_3$ that would induce a transition into the zig-zag ordered phase. 

\section{Thermal Pure Quantum States}

To capture thermal excitations of Eq. (1) in the main text, and examine how the Kitaev spin liquid is deformed in the presence of finite $\Gamma$, we systematically calculate heat capacity by using typical pure states\cite{PhysRevLett.99.160404}.
Several pioneering studies~\cite{JPSJ.55.3354,PhysRevB.49.5065,PhysRevE.62.4365} have shown that averages over a few tens of pure states can replace a canonical ensemble.
Recently, a statistical mechanics proof for the replacement of a canonical ensemble by a single pure state was given in Refs.\onlinecite{SSugiura2012,SSugiura2013}, where such a pure state that replaces a canonical ensemble is called a thermal pure quantum (TPQ) state\cite{SSugiura2012,SSugiura2013}.

We briefly summarize the construction of TPQ states following Ref.~\onlinecite{SSugiura2012}.
A TPQ state of $N$ quantum $S=1/2$ spins at infinite temperatures is simply given by a random vector, 
\begin{equation}
|\phi_{+\infty}\rangle=\sum_{i=0}^{2^N-1}c_i|i\rangle,
\end{equation}
where $|i\rangle$ is represented by the real-space $S=1/2$ basis and specified by a binary representation of a decimal number and $\{c_i\}$ is a set of random complex numbers with the normalization condition $\sum_{i=0}^{2^N-1} |c_i|^2=1$. 
Then, by multiplying a wave vector by the hamiltonian $\hat{H}$, we construct the TPQ states at lower temperatures as follows: Starting with an initial (or the 0-th step) vector $|\Phi_0\rangle=|\phi_{+\infty}\rangle$, the $k$-th step vector $|\Phi_k\rangle$ $(k\geq 1)$ is constructed as
\begin{equation}
|\Phi_k\rangle=\frac{(\Lambda-\hat{H})|\Phi_{k-1}\rangle}{\sqrt{\langle\Phi_{k-1}|(\Lambda-\hat{H})^2|\Phi_{k-1}\rangle}},
\end{equation}
where $\Lambda$ is a constant larger than the largest eigenvalue of $\hat{H}$.
The above $k$-th step vector is a TPQ state at a finite temperature $T$. 
The corresponding inverse temperature ${\beta}=(k_{B}T)^{-1}$ is determined through the following formula~\cite{SSugiura2012},
\begin{eqnarray}
\beta=\frac{2k_{B} k}{\Lambda-\langle\Phi_{k}|\hat{H}|\Phi_{k}\rangle}+O(1/N),
\end{eqnarray}
where $k_{B}$ is the Boltzmann constant.
In other word, a TPQ state at temperatures $T$ is given as,
\begin{eqnarray}
|\phi_{T}\rangle=|\Phi_k\rangle.
\end{eqnarray}

The heat capacity and entropy of $\hat{H}$ are then estimated by using a TPQ state $|\phi_{T}\rangle$.
In thermodynamic limit, a single TPQ state $|\phi_{T}\rangle$ indeed replaces a canonical ensemble and gives us exact heat capacity and entropy.
However, for finite $N$, contribution from atypical states is not negligible.
Therefore, in the present study, we estimate errors that originate from the atypical states by
using standard deviation of the results obtained from several initial random vectors.
Here, we calculate the heat capacity $C$ by using fluctuations of internal energy.
Ideally, the following equation holds:
\begin{eqnarray}
C(T)=k_{\rm B}\frac{\langle\phi_{T}|\hat{H}^2|\phi_{T}\rangle-\langle\phi_{T}|\hat{H}|\phi_{T}\rangle^2}{(k_{\rm B}T)^2}.
\label{def_C}
\end{eqnarray}
The entropy $S$ is estimated by integrating $C/T$ from high temperatures as
\begin{eqnarray}
S=Nk_{B}\ln2-\int_{T}^{+\infty}dT' \frac{C}{T'}.\label{def_S}
\end{eqnarray}

\section{\lowercase{i}TEBD method}

We also studied Eq. (1) in the main text on an infinite Cayley tree with $z=3$ connectivity, using the infinite time-evolving block decimation algorithm\cite{vidal_classical_2007} (iTEBD).
Within this approach, the ground state is approximated by a tensor product state (TPS) defined on an infinite tree lattice.
A Cayley tree TPS $|\psi\rangle$ may be defined by site-tensors $T_{i_rj_rk_r}^{(r),s_r}$, and diagonal bond-matrices $\lambda_{i_ri_{r'}}^{\langle rr'\rangle}$.
Here $r,r'$ denote sites, $s_r=\uparrow,\downarrow$ is a site's spin index, and $i_r,j_r,k_r$ are its bond indices, for the $x,y,z$ bonds, respectively, each running up to the bond dimension $\chi$.
The TPS is formally given by
\begin{eqnarray}
  \label{eq:TPS}
  |\psi\rangle & = & \prod_{r}\sum_{i_rj_rk_rs_r}T_{i_rj_rk_r}^{(r),s_r} \nonumber \\
  & & \times\prod_{\langle rr'\rangle\in x}\lambda_{i_ri_{r'}}^{\langle rr'\rangle}
  \prod_{\langle rr'\rangle\in y}\lambda_{j_rj_{r'}}^{\langle rr'\rangle}
  \prod_{\langle rr'\rangle\in z}\lambda_{k_rk_{r'}}^{\langle rr'\rangle}
  |s_r\rangle
\end{eqnarray}
In the TPS canonical form, the diagonal values of
$\lambda_{i_ri_{r'}}^{\langle rr'\rangle}$ determine the state's Schmidt
decomposition at the corresponding bond. Therefore, the entanglement
entropy for partitioning the system at that bond is simply given by
$S_E=-\sum_i\lambda_{ii}^2\log\lambda_{ii}^2$. Assuming the ground
sate is periodic, the Hilbert space can be greatly reduced by
considering only a subset of TPSs defined by a periodic arrangement of
a small number of independent tensors. In our calculation we consider
an eight-site periodicity which is
compatible with the simplest magnetically ordered states expected in
this system. Under such an assumption, there are eight independent
$T$ site-tensors and twelve independent $\lambda$ bond-matrices.

The ground state is obtained by projection, iteratively evolving the
TPS in imaginary time until convergence is reached. Technically, the
time evolution operator $U=e^{-\Delta\tau H}$, where $H$ is the
Hamiltonian, and $\Delta\tau$ is a small time step, is applied using a
fourth-order Suzuki-Trotter
decomposition\cite{sornborger_higher-order_1998}, separately evolving
each (independent) bond at a time. Each time the evolution operator is
applied, the bond dimension (or Schmidt rank) $\chi$ generally
increases, making the algorithm unmanageable. To avoid this, a fixed
$\chi$ is maintained by applying a truncated singular value
decomposition (SVD) to the evolved TPS. Thus, larger values of $\chi$
yield a more accurate ground state.  Additional pure SVD steps,
applied between time evolution operations, are needed in order to
bring the TPS closer to the canonical form\cite{orus_infinite_2008}.
We begin by iteratively evolving a random TPS with a time step of
$\Delta\tau=0.1$ until the entanglement entropies $S_E$ on all bonds
have converged. We then reduce $\Delta\tau$ by a half and continue
evolving, again until convergence is achieved. This process is
repeated until $\Delta\tau<10^{-5}$. Since this method is defined
directly in the thermodynamic limit, there are no finite size effects,
and one may arrive at ground states which spontaneously break
underlying symmetries.

Generally, TPSs are well suited to describe gapped phases, but not
gapless ones. In the Kitaev limit, it is possible to have a gapped
phase by introducing bond anisotropy\cite{kitaev_anyons_2006}.  When
applying the iTEBD method to Eq. (1), we similarly found that
the calculation was unstable in the isotropic limit, and a small bond
anisotropy was required to stabilize it.
By calculating the zero energy density of states for Majorana fermions hopping on the anisotropic infinite tree, Kimchi et al.\cite{kimchi_three-dimensional_2014} were able to determine that in the $K$ limit there is a gapped phase for $a_K\gtrsim0.05$.


\begin{thebibliography}{38}%
\makeatletter
\providecommand \@ifxundefined [1]{%
 \@ifx{#1\undefined}
}%
\providecommand \@ifnum [1]{%
 \ifnum #1\expandafter \@firstoftwo
 \else \expandafter \@secondoftwo
 \fi
}%
\providecommand \@ifx [1]{%
 \ifx #1\expandafter \@firstoftwo
 \else \expandafter \@secondoftwo
 \fi
}%
\providecommand \natexlab [1]{#1}%
\providecommand \enquote  [1]{``#1''}%
\providecommand \bibnamefont  [1]{#1}%
\providecommand \bibfnamefont [1]{#1}%
\providecommand \citenamefont [1]{#1}%
\providecommand \href@noop [0]{\@secondoftwo}%
\providecommand \href [0]{\begingroup \@sanitize@url \@href}%
\providecommand \@href[1]{\@@startlink{#1}\@@href}%
\providecommand \@@href[1]{\endgroup#1\@@endlink}%
\providecommand \@sanitize@url [0]{\catcode `\\12\catcode `\$12\catcode
  `\&12\catcode `\#12\catcode `\^12\catcode `\_12\catcode `\%12\relax}%
\providecommand \@@startlink[1]{}%
\providecommand \@@endlink[0]{}%
\providecommand \url  [0]{\begingroup\@sanitize@url \@url }%
\providecommand \@url [1]{\endgroup\@href {#1}{\urlprefix }}%
\providecommand \urlprefix  [0]{URL }%
\providecommand \Eprint [0]{\href }%
\providecommand \doibase [0]{http://dx.doi.org/}%
\providecommand \selectlanguage [0]{\@gobble}%
\providecommand \bibinfo  [0]{\@secondoftwo}%
\providecommand \bibfield  [0]{\@secondoftwo}%
\providecommand \translation [1]{[#1]}%
\providecommand \BibitemOpen [0]{}%
\providecommand \bibitemStop [0]{}%
\providecommand \bibitemNoStop [0]{.\EOS\space}%
\providecommand \EOS [0]{\spacefactor3000\relax}%
\providecommand \BibitemShut  [1]{\csname bibitem#1\endcsname}%
\let\auto@bib@innerbib\@empty
\bibitem [{\citenamefont {Balents}(2010)}]{Balents2010}%
  \BibitemOpen
  \bibfield  {author} {\bibinfo {author} {\bibfnamefont {L.}~\bibnamefont
  {Balents}},\ }\href@noop {} {\bibfield {title} {\bibinfo {title} {Spin liquids in frustrated magnets, }}}{\bibfield  {journal} {\bibinfo  {journal}
  {Nature}\ }\textbf {\bibinfo {volume} {464}},\ \bibinfo {pages} {199}
  (\bibinfo {year} {2010})}\BibitemShut {NoStop}%
\bibitem [{\citenamefont {Shimizu}\ \emph {et~al.}(2003)\citenamefont
  {Shimizu}, \citenamefont {Miyagawa}, \citenamefont {Kanoda}, \citenamefont
  {Maesato},\ and\ \citenamefont {Saito}}]{Shimizu2003}%
  \BibitemOpen
  \bibfield  {author} {\bibinfo {author} {\bibfnamefont {Y.}~\bibnamefont
  {Shimizu}}, \bibinfo {author} {\bibfnamefont {K.}~\bibnamefont {Miyagawa}},
  \bibinfo {author} {\bibfnamefont {K.}~\bibnamefont {Kanoda}}, \bibinfo
  {author} {\bibfnamefont {M.}~\bibnamefont {Maesato}}, \ and\ \bibinfo
  {author} {\bibfnamefont {G.}~\bibnamefont {Saito}},\ }\href@noop {}
  {\bibfield {title} {\bibinfo {title} {Spin Liquid State in an Organic Mott Insulator with a Triangular Lattice, }}}{\bibfield  {journal} {\bibinfo  {journal} {Phys. Rev. Lett.}\ }\textbf
  {\bibinfo {volume} {91}},\ \bibinfo {pages} {107001} (\bibinfo {year}
  {2003})}\BibitemShut {NoStop}%
\bibitem [{\citenamefont {Helton}\ \emph {et~al.}(2007)\citenamefont {Helton},
  \citenamefont {Matan}, \citenamefont {Shores}, \citenamefont {Nytko},
  \citenamefont {Bartlett}, \citenamefont {Yoshida}, \citenamefont {Takano},
  \citenamefont {Suslov}, \citenamefont {Qiu}, \citenamefont {Chung},
  \citenamefont {Nocera},\ and\ \citenamefont {Lee}}]{Helton2007}%
  \BibitemOpen
  \bibfield  {author} {\bibinfo {author} {\bibfnamefont {J.~S.}\ \bibnamefont
  {Helton}}, \bibinfo {author} {\bibfnamefont {K.}~\bibnamefont {Matan}},
  \bibinfo {author} {\bibfnamefont {M.~P.}\ \bibnamefont {Shores}}, \bibinfo
  {author} {\bibfnamefont {E.~A.}\ \bibnamefont {Nytko}}, \bibinfo {author}
  {\bibfnamefont {B.~M.}\ \bibnamefont {Bartlett}}, \bibinfo {author}
  {\bibfnamefont {Y.}~\bibnamefont {Yoshida}}, \bibinfo {author} {\bibfnamefont
  {Y.}~\bibnamefont {Takano}}, \bibinfo {author} {\bibfnamefont
  {A.}~\bibnamefont {Suslov}}, \bibinfo {author} {\bibfnamefont
  {Y.}~\bibnamefont {Qiu}}, \bibinfo {author} {\bibfnamefont {J.-H.}\
  \bibnamefont {Chung}}, \bibinfo {author} {\bibfnamefont {D.~G.}\ \bibnamefont
  {Nocera}}, \ and\ \bibinfo {author} {\bibfnamefont {Y.~S.}\ \bibnamefont
  {Lee}},\ }\href@noop {} {\bibfield {title} {\bibinfo {title} {Spin Dynamics of the Spin-$\frac{1}{2}$ Kagome Lattice Antiferromagnet ZnCu$_3$(OH)$_6$Cl$_2$, }}}{\bibfield  {journal} {\bibinfo  {journal} {Phys.
  Rev. Lett.}\ }\textbf {\bibinfo {volume} {98}},\ \bibinfo {pages} {107204}
  (\bibinfo {year} {2007})}\BibitemShut {NoStop}%
\bibitem [{\citenamefont {Okamoto}\ \emph {et~al.}(2007)\citenamefont
  {Okamoto}, \citenamefont {Nohara}, \citenamefont {Aruga-Katori},\ and\
  \citenamefont {Takagi}}]{Okamoto2007}%
  \BibitemOpen
  \bibfield  {author} {\bibinfo {author} {\bibfnamefont {Y.}~\bibnamefont
  {Okamoto}}, \bibinfo {author} {\bibfnamefont {M.}~\bibnamefont {Nohara}},
  \bibinfo {author} {\bibfnamefont {H.}~\bibnamefont {Aruga-Katori}}, \ and\
  \bibinfo {author} {\bibfnamefont {H.}~\bibnamefont {Takagi}},\ }\href@noop {}
  {\bibfield {title} {\bibinfo {title} {Spin-Liquid State in the S = $\frac{1}{2}$ Hyperkagome Antiferromagnet Na$_4$Ir$_3$O$_8$, }}}{\bibfield  {journal} {\bibinfo  {journal} {Phys. Rev. Lett.}\ }\textbf
  {\bibinfo {volume} {99}},\ \bibinfo {pages} {137207} (\bibinfo {year}
  {2007})}\BibitemShut {NoStop}%
\bibitem [{\citenamefont {Yamashita}\ \emph {et~al.}(2010)\citenamefont
  {Yamashita}, \citenamefont {Nakata}, \citenamefont {Senshu}, \citenamefont
  {Nagata}, \citenamefont {Yamamoto}, \citenamefont {Kato}, \citenamefont
  {Shibauchi},\ and\ \citenamefont {Matsuda}}]{MYamashita2010}%
  \BibitemOpen
  \bibfield  {author} {\bibinfo {author} {\bibfnamefont {M.}~\bibnamefont
  {Yamashita}}, \bibinfo {author} {\bibfnamefont {N.}~\bibnamefont {Nakata}},
  \bibinfo {author} {\bibfnamefont {Y.}~\bibnamefont {Senshu}}, \bibinfo
  {author} {\bibfnamefont {M.}~\bibnamefont {Nagata}}, \bibinfo {author}
  {\bibfnamefont {H.~M.}\ \bibnamefont {Yamamoto}}, \bibinfo {author}
  {\bibfnamefont {R.}~\bibnamefont {Kato}}, \bibinfo {author} {\bibfnamefont
  {T.}~\bibnamefont {Shibauchi}}, \ and\ \bibinfo {author} {\bibfnamefont
  {Y.}~\bibnamefont {Matsuda}},\ }\href@noop {} {\bibfield {title} {\bibinfo {title} {Highly Mobile Gapless Excitations in a Two-Dimensional Candidate Quantum Spin Liquid, }}}{\bibfield  {journal} {\bibinfo
   {journal} {Science}\ }\textbf {\bibinfo {volume} {328}},\ \bibinfo {pages}
  {1246} (\bibinfo {year} {2010})}\BibitemShut {NoStop}%
\bibitem [{\citenamefont {Han}\ \emph {et~al.}(2012)\citenamefont {Han},
  \citenamefont {Helton}, \citenamefont {Chu}, \citenamefont {Nocera},
  \citenamefont {Rodriguez-Rivera}, \citenamefont {Broholm},\ and\
  \citenamefont {Lee}}]{Han2012}%
  \BibitemOpen
  \bibfield  {author} {\bibinfo {author} {\bibfnamefont {T.-H.}\ \bibnamefont
  {Han}}, \bibinfo {author} {\bibfnamefont {J.~S.}\ \bibnamefont {Helton}},
  \bibinfo {author} {\bibfnamefont {S.}~\bibnamefont {Chu}}, \bibinfo {author}
  {\bibfnamefont {D.~G.}\ \bibnamefont {Nocera}}, \bibinfo {author}
  {\bibfnamefont {J.~A.}\ \bibnamefont {Rodriguez-Rivera}}, \bibinfo {author}
  {\bibfnamefont {C.}~\bibnamefont {Broholm}}, \ and\ \bibinfo {author}
  {\bibfnamefont {Y.~S.}\ \bibnamefont {Lee}},\ }\href@noop {} {\bibfield {title} {\bibinfo {title} {Fractionalized excitations in the spin-liquid state of a kagome-lattice antiferromagnet, }}}{\bibfield
  {journal} {\bibinfo  {journal} {Nature}\ }\textbf {\bibinfo {volume} {492}},\
  \bibinfo {pages} {406} (\bibinfo {year} {2012})}\BibitemShut {NoStop}%
\bibitem [{\citenamefont {Jackeli}\ and\ \citenamefont
  {Khaliullin}(2009)}]{Jackeli2009}%
  \BibitemOpen
  \bibfield  {author} {\bibinfo {author} {\bibfnamefont {G.}~\bibnamefont
  {Jackeli}}\ and\ \bibinfo {author} {\bibfnamefont {G.}~\bibnamefont
  {Khaliullin}},\ }\href {\doibase 10.1103/PhysRevLett.102.017205} {\bibfield {title} {\bibinfo {title} {Mott Insulators in the Strong Spin-Orbit Coupling Limit: From Heisenberg to a Quantum Compass and Kitaev Models, }}}{\bibfield
  {journal} {\bibinfo  {journal} {Phys. Rev. Lett.}\ }\textbf {\bibinfo
  {volume} {102}},\ \bibinfo {pages} {017205} (\bibinfo {year}
  {2009})}\BibitemShut {NoStop}%
\bibitem [{\citenamefont {Witczak-Krempa}\ \emph {et~al.}(2013)\citenamefont
  {Witczak-Krempa}, \citenamefont {Chen}, \citenamefont {Kim},\ and\
  \citenamefont {Balents}}]{WCKB2013}%
  \BibitemOpen
  \bibfield  {author} {\bibinfo {author} {\bibfnamefont {W.}~\bibnamefont
  {Witczak-Krempa}}, \bibinfo {author} {\bibfnamefont {G.}~\bibnamefont
  {Chen}}, \bibinfo {author} {\bibfnamefont {Y.~B.}\ \bibnamefont {Kim}}, \
  and\ \bibinfo {author} {\bibfnamefont {L.}~\bibnamefont {Balents}},\ }\href
  {\doibase 10.1146/annurev-conmatphys-020911-125138} {\bibfield {title} {\bibinfo {title} {Correlated Quantum Phenomena in the Strong Spin-Orbit Regime, }}}{\bibfield  {journal}
  {\bibinfo  {journal} {Annual Review of Condensed Matter Physics}\ }\textbf
  {\bibinfo {volume} {5}},\ \bibinfo {pages} {57} (\bibinfo {year}
  {2013})}\BibitemShut {NoStop}%
\bibitem [{\citenamefont {Rau}\ \emph {et~al.}(2016)\citenamefont {Rau},
  \citenamefont {Lee},\ and\ \citenamefont {Kee}}]{RLK2016}%
  \BibitemOpen
  \bibfield  {author} {\bibinfo {author} {\bibfnamefont {J.~G.}\ \bibnamefont
  {Rau}}, \bibinfo {author} {\bibfnamefont {E.~K.-H.}\ \bibnamefont {Lee}}, \
  and\ \bibinfo {author} {\bibfnamefont {H.-Y.}\ \bibnamefont {Kee}},\ }\href
  {\doibase 10.1146/annurev-conmatphys-031115-011319} {\bibfield {title} {\bibinfo {title} {Spin-Orbit Physics Giving Rise to Novel Phases in Correlated Systems: Iridates and Related Materials, }}}{\bibfield  {journal}
  {\bibinfo  {journal} {Annual Review of Condensed Matter Physics}\ }\textbf
  {\bibinfo {volume} {7}},\ \bibinfo {pages} {195} (\bibinfo {year}
  {2016})}\BibitemShut {NoStop}%
\bibitem [{\citenamefont {Singh}\ \emph {et~al.}(2012)\citenamefont {Singh},
  \citenamefont {Manni}, \citenamefont {Reuther}, \citenamefont {Berlijn},
  \citenamefont {Thomale}, \citenamefont {Ku}, \citenamefont {Trebst},\ and\
  \citenamefont {Gegenwart}}]{Singh2012}%
  \BibitemOpen
  \bibfield  {author} {\bibinfo {author} {\bibfnamefont {Y.}~\bibnamefont
  {Singh}}, \bibinfo {author} {\bibfnamefont {S.}~\bibnamefont {Manni}},
  \bibinfo {author} {\bibfnamefont {J.}~\bibnamefont {Reuther}}, \bibinfo
  {author} {\bibfnamefont {T.}~\bibnamefont {Berlijn}}, \bibinfo {author}
  {\bibfnamefont {R.}~\bibnamefont {Thomale}}, \bibinfo {author} {\bibfnamefont
  {W.}~\bibnamefont {Ku}}, \bibinfo {author} {\bibfnamefont {S.}~\bibnamefont
  {Trebst}}, \ and\ \bibinfo {author} {\bibfnamefont {P.}~\bibnamefont
  {Gegenwart}},\ }\href {\doibase 10.1103/PhysRevLett.108.127203} {\bibfield {title} {\bibinfo {title} {Relevance of the Heisenberg-Kitaev Model for the Honeycomb Lattice Iridates A$_2$IrO$_3$, }}}{\bibfield
  {journal} {\bibinfo  {journal} {Phys. Rev. Lett.}\ }\textbf {\bibinfo
  {volume} {108}},\ \bibinfo {pages} {127203} (\bibinfo {year}
  {2012})}\BibitemShut {NoStop}%
\bibitem [{\citenamefont {Plumb}\ \emph {et~al.}(2014)\citenamefont {Plumb},
  \citenamefont {Clancy}, \citenamefont {Sandilands}, \citenamefont
  {Vijay~Shankar}, \citenamefont {Hu}, \citenamefont {Burch}, \citenamefont
  {Kee},\ and\ \citenamefont {Kim}}]{Plumb2014}%
  \BibitemOpen
  \bibfield  {author} {\bibinfo {author} {\bibfnamefont {K.}~\bibnamefont
  {Plumb}}, \bibinfo {author} {\bibfnamefont {J.}~\bibnamefont {Clancy}},
  \bibinfo {author} {\bibfnamefont {L.}~\bibnamefont {Sandilands}}, \bibinfo
  {author} {\bibfnamefont {V.}~\bibnamefont {Vijay~Shankar}}, \bibinfo {author}
  {\bibfnamefont {Y.}~\bibnamefont {Hu}}, \bibinfo {author} {\bibfnamefont
  {K.}~\bibnamefont {Burch}}, \bibinfo {author} {\bibfnamefont {H.-Y.}\
  \bibnamefont {Kee}}, \ and\ \bibinfo {author} {\bibfnamefont {Y.-J.}\
  \bibnamefont {Kim}},\ }\href {\doibase 10.1103/PhysRevB.90.041112} {\bibfield {title} {\bibinfo {title} {$\alpha$-RuCl$_3$: A spin-orbit assisted Mott insulator on a honeycomb lattice, }}}{\bibfield
   {journal} {\bibinfo  {journal} {Phys. Rev. B}\ }\textbf {\bibinfo {volume}
  {90}},\ \bibinfo {pages} {041112} (\bibinfo {year} {2014})}\BibitemShut
  {NoStop}%
\bibitem [{\citenamefont {Modic}\ \emph {et~al.}(2014)\citenamefont {Modic},
  \citenamefont {Smidt}, \citenamefont {Kimchi}, \citenamefont {Breznay},
  \citenamefont {Biffin}, \citenamefont {Choi}, \citenamefont {Johnson},
  \citenamefont {Coldea}, \citenamefont {Watkins-Curry}, \citenamefont
  {McCandless}, \citenamefont {Chan}, \citenamefont {Gandara}, \citenamefont
  {Islam}, \citenamefont {Vishwanath}, \citenamefont {Shekhter}, \citenamefont
  {McDonald},\ and\ \citenamefont {Analytis}}]{Modic2014}%
  \BibitemOpen
  \bibfield  {author} {\bibinfo {author} {\bibfnamefont {K.~A.}\ \bibnamefont
  {Modic}}, \bibinfo {author} {\bibfnamefont {T.~E.}\ \bibnamefont {Smidt}},
  \bibinfo {author} {\bibfnamefont {I.}~\bibnamefont {Kimchi}}, \bibinfo
  {author} {\bibfnamefont {N.~P.}\ \bibnamefont {Breznay}}, \bibinfo {author}
  {\bibfnamefont {A.}~\bibnamefont {Biffin}}, \bibinfo {author} {\bibfnamefont
  {S.}~\bibnamefont {Choi}}, \bibinfo {author} {\bibfnamefont {R.~D.}\
  \bibnamefont {Johnson}}, \bibinfo {author} {\bibfnamefont {R.}~\bibnamefont
  {Coldea}}, \bibinfo {author} {\bibfnamefont {P.}~\bibnamefont
  {Watkins-Curry}}, \bibinfo {author} {\bibfnamefont {G.~T.}\ \bibnamefont
  {McCandless}}, \bibinfo {author} {\bibfnamefont {J.~Y.}\ \bibnamefont
  {Chan}}, \bibinfo {author} {\bibfnamefont {F.}~\bibnamefont {Gandara}},
  \bibinfo {author} {\bibfnamefont {Z.}~\bibnamefont {Islam}}, \bibinfo
  {author} {\bibfnamefont {A.}~\bibnamefont {Vishwanath}}, \bibinfo {author}
  {\bibfnamefont {A.}~\bibnamefont {Shekhter}}, \bibinfo {author}
  {\bibfnamefont {R.~D.}\ \bibnamefont {McDonald}}, \ and\ \bibinfo {author}
  {\bibfnamefont {J.~G.}\ \bibnamefont {Analytis}},\ }\href@noop {} {\bibfield {title} {\bibinfo {title} {Realization of a three-dimensional spin–anisotropic harmonic honeycomb iridate, }}}{\bibfield
  {journal} {\bibinfo  {journal} {Nature Communications}\ }\textbf {\bibinfo
  {volume} {5}},\ \bibinfo {pages} {4203} (\bibinfo {year} {2014})}\BibitemShut
  {NoStop}%
\bibitem [{\citenamefont {Takayama}\ \emph {et~al.}(2015)\citenamefont
  {Takayama}, \citenamefont {Kato}, \citenamefont {Dinnebier}, \citenamefont
  {Nuss}, \citenamefont {Kono}, \citenamefont {Veiga}, \citenamefont {Fabbris},
  \citenamefont {Haskel},\ and\ \citenamefont {Takagi}}]{Takayama2015}%
  \BibitemOpen
  \bibfield  {author} {\bibinfo {author} {\bibfnamefont {T.}~\bibnamefont
  {Takayama}}, \bibinfo {author} {\bibfnamefont {A.}~\bibnamefont {Kato}},
  \bibinfo {author} {\bibfnamefont {R.}~\bibnamefont {Dinnebier}}, \bibinfo
  {author} {\bibfnamefont {J.}~\bibnamefont {Nuss}}, \bibinfo {author}
  {\bibfnamefont {H.}~\bibnamefont {Kono}}, \bibinfo {author} {\bibfnamefont
  {L.}~\bibnamefont {Veiga}}, \bibinfo {author} {\bibfnamefont
  {G.}~\bibnamefont {Fabbris}}, \bibinfo {author} {\bibfnamefont
  {D.}~\bibnamefont {Haskel}}, \ and\ \bibinfo {author} {\bibfnamefont
  {H.}~\bibnamefont {Takagi}},\ }\href@noop {} {\bibfield {title} {\bibinfo {title} {Hyperhoneycomb Iridate $\beta$-Li$_2$IrO$_3$ as a Platform for Kitaev Magnetism, }}}{\bibfield  {journal} {\bibinfo
  {journal} {Phys. Rev. Lett.}\ }\textbf {\bibinfo {volume} {114}},\ \bibinfo
  {pages} {077202} (\bibinfo {year} {2015})}\BibitemShut {NoStop}%
\bibitem [{\citenamefont {Rau}\ \emph {et~al.}(2014)\citenamefont {Rau},
  \citenamefont {Lee},\ and\ \citenamefont {Kee}}]{Rau2014}%
  \BibitemOpen
  \bibfield  {author} {\bibinfo {author} {\bibfnamefont {J.~G.}\ \bibnamefont
  {Rau}}, \bibinfo {author} {\bibfnamefont {E.~K.-H.}\ \bibnamefont {Lee}}, \
  and\ \bibinfo {author} {\bibfnamefont {H.-Y.}\ \bibnamefont {Kee}},\ }\href
  {\doibase 10.1103/PhysRevLett.112.077204} {\bibfield {title} {\bibinfo {title} {Generic Spin Model for the Honeycomb Iridates beyond the Kitaev Limit, }}}{\bibfield  {journal} {\bibinfo
  {journal} {Phys. Rev. Lett.}\ }\textbf {\bibinfo {volume} {112}},\ \bibinfo
  {pages} {077204} (\bibinfo {year} {2014})}\BibitemShut {NoStop}%
\bibitem [{\citenamefont {Ran}\ \emph {et~al.}(2017)}]{Ran2017}%
  \BibitemOpen
  \bibfield  {author} {\bibinfo {author} {\bibfnamefont {K.}\ \bibnamefont
      {Ran}}, \
    \bibinfo {author} {\bibfnamefont {J.}\ \bibnamefont {Wang}},\
    \bibinfo {author} {\bibfnamefont {W.}\ \bibnamefont {Wang}},\
    \bibinfo {author} {\bibfnamefont {Z.-Y.}\ \bibnamefont {Dong}},\
    \bibinfo {author} {\bibfnamefont {X.}\ \bibnamefont {Ren}},\
    \bibinfo {author} {\bibfnamefont {S.}\ \bibnamefont {Bao}},\
    \bibinfo {author} {\bibfnamefont {S.}\ \bibnamefont {Li}},\
    \bibinfo {author} {\bibfnamefont {Z.}\ \bibnamefont {Ma}},\
    \bibinfo {author} {\bibfnamefont {Y.}\ \bibnamefont {Gan}},\
    \bibinfo {author} {\bibfnamefont {Y.}\ \bibnamefont {Zhang}},\
    \bibinfo {author} {\bibfnamefont {J. T.}\ \bibnamefont {Park}},\
    \bibinfo {author} {\bibfnamefont {G.}\ \bibnamefont {Deng}},\
    \bibinfo {author} {\bibfnamefont {S.}\ \bibnamefont {Danilkin}},\
    \bibinfo {author} {\bibfnamefont {S.-L.}\ \bibnamefont {Yu}},\
    \bibinfo {author} {\bibfnamefont {J.-X.}\ \bibnamefont {Li}},\
    \bibinfo {author} {\bibfnamefont {J.}\ \bibnamefont {Wen}},\
  }\href
  {\doibase 10.1103/PhysRevLett.118.107203} {\bibfield {title} {\bibinfo {title} {Spin-Wave Excitations Evidencing the Kitaev Interaction in Single Crystalline $\alpha$-RuCl$_3$, }}}{\bibfield  {journal} {\bibinfo
  {journal} {Phys. Rev. Lett.}\ }\textbf {\bibinfo {volume} {118}},\ \bibinfo
  {pages} {107203} (\bibinfo {year} {2017})}\BibitemShut {NoStop}%
\bibitem [{\citenamefont {Wang}\ \emph {et~al.}(2017)}]{Wang2017}%
  \BibitemOpen
  \bibfield  {author} {\bibinfo {author} {\bibfnamefont {W.}\ \bibnamefont
      {Wang}}, \
    \bibinfo {author} {\bibfnamefont {Z.-Y.}\ \bibnamefont {Dong}},\
    \bibinfo {author} {\bibfnamefont {S.-L.}\ \bibnamefont {Yu}},\
    \bibinfo {author} {\bibfnamefont {J.-X.}\ \bibnamefont {Li}},\
  }\href
  {\doibase 10.1103/PhysRevB.96.115103} {\bibfield {title} {\bibinfo {title} {Theoretical investigation of magnetic dynamics in $\alpha$-RuCl$_3$, }}}{\bibfield  {journal} {\bibinfo
  {journal} {Phys. Rev. B}\ }\textbf {\bibinfo {volume} {96}},\ \bibinfo
  {pages} {115103} (\bibinfo {year} {2017})}\BibitemShut {NoStop}%
\bibitem [{\citenamefont {Kim}\ and\ \citenamefont {Kee}(2016)}]{Kim2016}%
  \BibitemOpen
  \bibfield  {author} {\bibinfo {author} {\bibfnamefont {H.-S.}\ \bibnamefont
  {Kim}}\ and\ \bibinfo {author} {\bibfnamefont {H.-Y.}\ \bibnamefont {Kee}},\
  }\href {\doibase 10.1103/PhysRevB.93.155143} {\bibfield {title} {\bibinfo {title} {Crystal structure and magnetism in $\alpha$-RuCl$_3$: An {\it ab initio} study, }}}{\bibfield  {journal} {\bibinfo
  {journal} {Phys. Rev. B}\ }\textbf {\bibinfo {volume} {93}},\ \bibinfo
  {pages} {155143} (\bibinfo {year} {2016})}\BibitemShut {NoStop}%
\bibitem [{\citenamefont {Winter}\ \emph {et~al.}(2016)\citenamefont {Winter},
  \citenamefont {Li}, \citenamefont {Jeschke},\ and\ \citenamefont
  {Valenti}}]{Winter2016}%
  \BibitemOpen
  \bibfield  {author} {\bibinfo {author} {\bibfnamefont {S.~M.}\ \bibnamefont
  {Winter}}, \bibinfo {author} {\bibfnamefont {Y.}~\bibnamefont {Li}}, \bibinfo
  {author} {\bibfnamefont {H.~O.}\ \bibnamefont {Jeschke}}, \ and\ \bibinfo
  {author} {\bibfnamefont {R.}~\bibnamefont {Valenti}},\ }\href@noop {}
  {\bibfield {title} {\bibinfo {title} {Challenges in design of Kitaev materials: Magnetic interactions from competing energy scales, }}}{\bibfield  {journal} {\bibinfo  {journal} {Phys. Rev. B}\ }\textbf {\bibinfo
  {volume} {93}},\ \bibinfo {pages} {214431} (\bibinfo {year}
  {2016})}\BibitemShut {NoStop}%
\bibitem [{\citenamefont {Yadav}\ \emph {et~al.}(2016)\citenamefont {Yadav},
  \citenamefont {Bogdanov}, \citenamefont {Katukuri}, \citenamefont
  {Nishimoto}, \citenamefont {van~den Brink},\ and\ \citenamefont
  {Hozoic}}]{Yadav2016}%
  \BibitemOpen
  \bibfield  {author} {\bibinfo {author} {\bibfnamefont {R.}~\bibnamefont
  {Yadav}}, \bibinfo {author} {\bibfnamefont {N.~A.}\ \bibnamefont {Bogdanov}},
  \bibinfo {author} {\bibfnamefont {V.~M.}\ \bibnamefont {Katukuri}}, \bibinfo
  {author} {\bibfnamefont {S.}~\bibnamefont {Nishimoto}}, \bibinfo {author}
  {\bibfnamefont {J.}~\bibnamefont {van~den Brink}}, \ and\ \bibinfo {author}
  {\bibfnamefont {L.}~\bibnamefont {Hozoic}},\ }\href@noop {} {\bibfield {title} {\bibinfo {title} {Kitaev exchange and field-induced quantum spin-liquid states in honeycomb $\alpha$-RuCl$_3$, }}}{\bibfield
  {journal} {\bibinfo  {journal} {Scientific Reports}\ }\textbf {\bibinfo
  {volume} {6}},\ \bibinfo {pages} {37925} (\bibinfo {year}
  {2016})}\BibitemShut {NoStop}%
\bibitem [{\citenamefont {Rau}\ and\ \citenamefont {Kee}(2014)}]{RauKeeTD}%
  \BibitemOpen
  \bibfield  {author} {\bibinfo {author} {\bibfnamefont {J.~G.}\ \bibnamefont
  {Rau}}\ and\ \bibinfo {author} {\bibfnamefont {H.-Y.}\ \bibnamefont {Kee}},\
  }\href {https://arxiv.org/abs/1408.4811} {\bibfield {title} {\bibinfo {title} {Trigonal distortion in the honeycomb iridates: Proximity of zigzag and spiral phases in Na$_2$IrO$_3$, }}}{\bibfield  {journal} {\bibinfo
  {journal} {arXiv preprint arXiv:1408.4811}\ } (\bibinfo {year}
  {2014})}\BibitemShut {NoStop}%
\bibitem [{\citenamefont {Rousochatzakis}\ and\ \citenamefont
  {Perkins}(2017)}]{rousochatzakis_classical_2016}%
  \BibitemOpen
  \bibfield  {author} {\bibinfo {author} {\bibfnamefont {I.}~\bibnamefont
  {Rousochatzakis}}\ and\ \bibinfo {author} {\bibfnamefont {N. B.}~\bibnamefont
  {Perkins}},\ }\href@noop {} {\bibfield {title} {\bibinfo {title} {Classical Spin Liquid Instability Driven By Off-Diagonal Exchange in Strong Spin-Orbit Magnets, }}}{\bibfield  {journal} {\bibinfo  {journal}
  {Phys. Rev. Lett.}\ }\textbf {\bibinfo {volume} {118}},\ \bibinfo {pages} {147204}
  (\bibinfo {year} {2018})}\BibitemShut {NoStop}%
\bibitem [{\citenamefont {Gohlke}(2017)}]{Gohlke2017}%
  \BibitemOpen
  \bibfield  {author} {\bibinfo {author} {\bibfnamefont {M.}\ \bibnamefont
      {Gohlke}},\ \bibinfo {author} {\bibfnamefont {G.}\ \bibnamefont {Wachtel}},\
    \bibinfo {author} {\bibfnamefont {Y.}\ \bibnamefont {Yamaji}},\
    \bibinfo {author} {\bibfnamefont {F.}\ \bibnamefont {Pollmann}},\
    \bibinfo {author} {\bibfnamefont {Y. B.}\ \bibnamefont {Kim}},\    
  }\href {} {\bibfield {title} {\bibinfo {title} {Quantum spin liquid signatures in Kitaev-like frustrated magnets, }}}{\bibfield  {journal} {\bibinfo
  {journal} {Phys. Rev. B}\ } \textbf {\bibinfo {volume} {97}},\ \bibinfo {pages} {075126}
  (\bibinfo {year} {2018})}\BibitemShut {NoStop}%
\bibitem [{\citenamefont {Imada}\ and\ \citenamefont
  {Takahashi}(1986)}]{JPSJ.55.3354}%
  \BibitemOpen
  \bibfield  {author} {\bibinfo {author} {\bibfnamefont {M.}~\bibnamefont
  {Imada}}\ and\ \bibinfo {author} {\bibfnamefont {M.}~\bibnamefont
  {Takahashi}},\ }\href {\doibase 10.1143/JPSJ.55.3354} {\bibfield {title} {\bibinfo {title} {Quantum Transfer Monte Carlo Method for Finite Temperature Properties and Quantum Molecular Dynamics Method for Dynamical Correlation Functions, }}}{\bibfield  {journal}
  {\bibinfo  {journal} {J. Phys. Soc. Jpn.}\ }\textbf {\bibinfo {volume}
  {55}},\ \bibinfo {pages} {3354} (\bibinfo {year} {1986})}\BibitemShut
  {NoStop}%
\bibitem [{\citenamefont {Jaklic}\ and\ \citenamefont
  {Prelovsek}(1994)}]{PhysRevB.49.5065}%
  \BibitemOpen
  \bibfield  {author} {\bibinfo {author} {\bibfnamefont {J.}~\bibnamefont
  {Jaklic}}\ and\ \bibinfo {author} {\bibfnamefont {P.}~\bibnamefont
  {Prelovsek}},\ }\href@noop {} {\bibfield {title} {\bibinfo {title} {Lanczos method for the calculation of finite-temperature quantities in correlated systems, }}}{\bibfield  {journal} {\bibinfo  {journal}
  {Phys. Rev. B}\ }\textbf {\bibinfo {volume} {49}},\ \bibinfo {pages} {5065}
  (\bibinfo {year} {1994})}\BibitemShut {NoStop}%
\bibitem [{\citenamefont {Hams}\ and\ \citenamefont
  {De~Raedt}(2000)}]{PhysRevE.62.4365}%
  \BibitemOpen
  \bibfield  {author} {\bibinfo {author} {\bibfnamefont {A.}~\bibnamefont
  {Hams}}\ and\ \bibinfo {author} {\bibfnamefont {H.}~\bibnamefont
  {De~Raedt}},\ }\href@noop {} {\bibfield {title} {\bibinfo {title} {Fast algorithm for finding the eigenvalue distribution of very large matrices, }}}{\bibfield  {journal} {\bibinfo  {journal}
  {Phys. Rev. E}\ }\textbf {\bibinfo {volume} {62}},\ \bibinfo {pages} {4365}
  (\bibinfo {year} {2000})}\BibitemShut {NoStop}%
\bibitem [{\citenamefont {Sugiura}\ and\ \citenamefont
  {Shimizu}(2012)}]{SSugiura2012}%
  \BibitemOpen
  \bibfield  {author} {\bibinfo {author} {\bibfnamefont {S.}~\bibnamefont
  {Sugiura}}\ and\ \bibinfo {author} {\bibfnamefont {A.}~\bibnamefont
  {Shimizu}},\ }\href@noop {} {\bibfield {title} {\bibinfo {title} {Thermal Pure Quantum States at Finite Temperature, }}}{\bibfield  {journal} {\bibinfo  {journal} {Phys.  Rev. Lett.}\ }\textbf {\bibinfo {volume} {108}},\ \bibinfo {pages} {240401}
  (\bibinfo {year} {2012})}\BibitemShut {NoStop}%
\bibitem [{\citenamefont {Sugiura}\ and\ \citenamefont
  {Shimizu}(2013)}]{SSugiura2013}%
  \BibitemOpen
  \bibfield  {author} {\bibinfo {author} {\bibfnamefont {S.}~\bibnamefont
  {Sugiura}}\ and\ \bibinfo {author} {\bibfnamefont {A.}~\bibnamefont
  {Shimizu}},\ }\href@noop {} {\bibfield {title} {\bibinfo {title} {Canonical Thermal Pure Quantum State, }}}{\bibfield  {journal} {\bibinfo  {journal} {Phys.
  Rev. Lett.}\ }\textbf {\bibinfo {volume} {111}},\ \bibinfo {pages} {010401}
  (\bibinfo {year} {2013})}\BibitemShut {NoStop}%
\bibitem [{\citenamefont {Chaloupka}\ and\ \citenamefont
  {Khaliullin}(2015)}]{Chaloupka2015}%
  \BibitemOpen
  \bibfield  {author} {\bibinfo {author} {\bibfnamefont {J.}~\bibnamefont
  {Chaloupka}}\ and\ \bibinfo {author} {\bibfnamefont {G.}~\bibnamefont
  {Khaliullin}},\ }\href@noop {} {\bibfield {title} {\bibinfo {title} {Hidden symmetries of the extended Kitaev-Heisenberg model: Implications for the honeycomb-lattice iridates A$_2$IrO$_3$, }}}{\bibfield  {journal} {\bibinfo  {journal}
  {Phys. Rev. B}\ }\textbf {\bibinfo {volume} {92}},\ \bibinfo {pages} {024413}
  (\bibinfo {year} {2015})}\BibitemShut {NoStop}%
\bibitem [{\citenamefont {Sears}\ \emph {et~al.}(2015)\citenamefont {Sears},
  \citenamefont {Songvilay}, \citenamefont {Plumb}, \citenamefont {Clancy},
  \citenamefont {Qiu}, \citenamefont {Zhao}, \citenamefont {Parshall},\ and\
  \citenamefont {Kim}}]{Jennifer2015}%
  \BibitemOpen
  \bibfield  {author} {\bibinfo {author} {\bibfnamefont {J.~A.}\ \bibnamefont
  {Sears}}, \bibinfo {author} {\bibfnamefont {M.}~\bibnamefont {Songvilay}},
  \bibinfo {author} {\bibfnamefont {K.~W.}\ \bibnamefont {Plumb}}, \bibinfo
  {author} {\bibfnamefont {J.~P.}\ \bibnamefont {Clancy}}, \bibinfo {author}
  {\bibfnamefont {Y.}~\bibnamefont {Qiu}}, \bibinfo {author} {\bibfnamefont
  {Y.}~\bibnamefont {Zhao}}, \bibinfo {author} {\bibfnamefont {D.}~\bibnamefont
  {Parshall}}, \ and\ \bibinfo {author} {\bibfnamefont {Y.-J.}\ \bibnamefont
  {Kim}},\ }\href {\doibase 10.1103/PhysRevB.91.144420} {\bibfield {title} {\bibinfo {title} {Magnetic order in $\alpha$-RuCl$_3$: A honeycomb-lattice quantum magnet with strong spin-orbit coupling, }}}{\bibfield  {journal}
  {\bibinfo  {journal} {Phys. Rev. B}\ }\textbf {\bibinfo {volume} {91}},\
  \bibinfo {pages} {144420} (\bibinfo {year} {2015})}\BibitemShut {NoStop}%
\bibitem [{\citenamefont {Johnson}\ \emph {et~al.}(2015)\citenamefont
  {Johnson}, \citenamefont {Williams}, \citenamefont {Haghighirad},
  \citenamefont {Singleton}, \citenamefont {Zapf}, \citenamefont {Manuel},
  \citenamefont {Mazin}, \citenamefont {Li}, \citenamefont {Jeschke},
  \citenamefont {Valent\'{\i}},\ and\ \citenamefont {Coldea}}]{Johnson2015}%
  \BibitemOpen
  \bibfield  {author} {\bibinfo {author} {\bibfnamefont {R.~D.}\ \bibnamefont
  {Johnson}}, \bibinfo {author} {\bibfnamefont {S.~C.}\ \bibnamefont
  {Williams}}, \bibinfo {author} {\bibfnamefont {A.~A.}\ \bibnamefont
  {Haghighirad}}, \bibinfo {author} {\bibfnamefont {J.}~\bibnamefont
  {Singleton}}, \bibinfo {author} {\bibfnamefont {V.}~\bibnamefont {Zapf}},
  \bibinfo {author} {\bibfnamefont {P.}~\bibnamefont {Manuel}}, \bibinfo
  {author} {\bibfnamefont {I.~I.}\ \bibnamefont {Mazin}}, \bibinfo {author}
  {\bibfnamefont {Y.}~\bibnamefont {Li}}, \bibinfo {author} {\bibfnamefont
  {H.~O.}\ \bibnamefont {Jeschke}}, \bibinfo {author} {\bibfnamefont
  {R.}~\bibnamefont {Valent\'{\i}}}, \ and\ \bibinfo {author} {\bibfnamefont
  {R.}~\bibnamefont {Coldea}},\ }\href {\doibase 10.1103/PhysRevB.92.235119}
  {\bibfield {title} {\bibinfo {title} {Monoclinic crystal structure of $\alpha$-RuCl$_3$ and the zigzag antiferromagnetic ground state, }}}{\bibfield  {journal} {\bibinfo  {journal} {Phys. Rev. B}\ }\textbf {\bibinfo
  {volume} {92}},\ \bibinfo {pages} {235119} (\bibinfo {year}
  {2015})}\BibitemShut {NoStop}%
\bibitem [{\citenamefont {Cao}\ \emph {et~al.}(2016)\citenamefont {Cao},
  \citenamefont {Banerjee}, \citenamefont {Yan}, \citenamefont {Bridges},
  \citenamefont {Lumsden}, \citenamefont {Mandrus}, \citenamefont {Tennant},
  \citenamefont {Chakoumakos},\ and\ \citenamefont {Nagler}}]{Cao2016}%
  \BibitemOpen
  \bibfield  {author} {\bibinfo {author} {\bibfnamefont {H.}~\bibnamefont
  {Cao}}, \bibinfo {author} {\bibfnamefont {A.}~\bibnamefont {Banerjee}},
  \bibinfo {author} {\bibfnamefont {J.-Q.}\ \bibnamefont {Yan}}, \bibinfo
  {author} {\bibfnamefont {C.}~\bibnamefont {Bridges}}, \bibinfo {author}
  {\bibfnamefont {M.}~\bibnamefont {Lumsden}}, \bibinfo {author} {\bibfnamefont
  {D.}~\bibnamefont {Mandrus}}, \bibinfo {author} {\bibfnamefont
  {D.}~\bibnamefont {Tennant}}, \bibinfo {author} {\bibfnamefont
  {B.}~\bibnamefont {Chakoumakos}}, \ and\ \bibinfo {author} {\bibfnamefont
  {S.}~\bibnamefont {Nagler}},\ }\href@noop {} {\bibfield {title} {\bibinfo {title} {Low-temperature crystal and magnetic structure of $\alpha$-RuCl$_3$, }}}{\bibfield  {journal} {\bibinfo
  {journal} {Phys. Rev. B}\ }\textbf {\bibinfo {volume} {93}},\ \bibinfo
  {pages} {134423} (\bibinfo {year} {2016})}\BibitemShut {NoStop}%
\bibitem [{\citenamefont {Nasu}\ \emph {et~al.}(2015)\citenamefont {Nasu},
  \citenamefont {Udagama},\ and\ \citenamefont {Motome}}]{JNasu2015}%
  \BibitemOpen
  \bibfield  {author} {\bibinfo {author} {\bibfnamefont {J.}~\bibnamefont
  {Nasu}}, \bibinfo {author} {\bibfnamefont {M.}~\bibnamefont {Udagama}}, \
  and\ \bibinfo {author} {\bibfnamefont {Y.}~\bibnamefont {Motome}},\
  }\href@noop {} {\bibfield {title} {\bibinfo {title} {Thermal fractionalization of quantum spins in a Kitaev model: Temperature-linear specific heat and coherent transport of Majorana fermions, }}}{\bibfield  {journal} {\bibinfo  {journal} {Phys. Rev. B}\
  }\textbf {\bibinfo {volume} {92}},\ \bibinfo {pages} {115122} (\bibinfo
  {year} {2015})}\BibitemShut {NoStop}%
\bibitem [{\citenamefont {Yamaji}\ \emph {et~al.}(2016)\citenamefont {Yamaji},
  \citenamefont {Suzuki}, \citenamefont {Yamada}, \citenamefont {Suga},
  \citenamefont {Kawashima},\ and\ \citenamefont {Imada}}]{PhysRevB.93.174425}%
  \BibitemOpen
  \bibfield  {author} {\bibinfo {author} {\bibfnamefont {Y.}~\bibnamefont
  {Yamaji}}, \bibinfo {author} {\bibfnamefont {T.}~\bibnamefont {Suzuki}},
  \bibinfo {author} {\bibfnamefont {T.}~\bibnamefont {Yamada}}, \bibinfo
  {author} {\bibfnamefont {S.-i.}\ \bibnamefont {Suga}}, \bibinfo {author}
  {\bibfnamefont {N.}~\bibnamefont {Kawashima}}, \ and\ \bibinfo {author}
  {\bibfnamefont {M.}~\bibnamefont {Imada}},\ }\href@noop {} {\bibfield {title} {\bibinfo {title} {Clues and criteria for designing a Kitaev spin liquid revealed by thermal and spin excitations of the honeycomb iridate Na$_2$IrO$_3$, }}}{\bibfield
  {journal} {\bibinfo  {journal} {Phys. Rev. B}\ }\textbf {\bibinfo {volume}
  {93}},\ \bibinfo {pages} {174425} (\bibinfo {year} {2016})}\BibitemShut
  {NoStop}%
\bibitem [{Not()}]{Note1}%
  \BibitemOpen
  \href@noop {} {}\bibinfo {note} {The height of the plateau in the temperature
  dependence of entropy is also examined by using a 32 site
  cluster.}\BibitemShut {Stop}%
\bibitem [{\citenamefont {Kitaev}(2006)}]{Kitaev2006}%
  \BibitemOpen
  \bibfield  {author} {\bibinfo {author} {\bibfnamefont {A.}~\bibnamefont
  {Kitaev}},\ }\href {\doibase 10.1016/j.aop.2005.10.005} {\bibfield {title} {\bibinfo {title} {Anyons in an exactly solved model and beyond, }}}{\bibfield  {journal}
  {\bibinfo  {journal} {Ann. Phys.}\ }\textbf {\bibinfo {volume} {321}},\
  \bibinfo {pages} {2} (\bibinfo {year} {2006})}\BibitemShut {NoStop}%
\bibitem [{\citenamefont {Knolle}\ \emph {et~al.}(2014)\citenamefont {Knolle},
  \citenamefont {Kovrizhin}, \citenamefont {Chalker},\ and\ \citenamefont
  {Moessner}}]{Knolle2014}%
  \BibitemOpen
  \bibfield  {author} {\bibinfo {author} {\bibfnamefont {J.}~\bibnamefont
  {Knolle}}, \bibinfo {author} {\bibfnamefont {D.}~\bibnamefont {Kovrizhin}},
  \bibinfo {author} {\bibfnamefont {J.}~\bibnamefont {Chalker}}, \ and\
  \bibinfo {author} {\bibfnamefont {R.}~\bibnamefont {Moessner}},\ }\href@noop
  {} {\bibfield {title} {\bibinfo {title} {Dynamics of a Two-Dimensional Quantum Spin Liquid: Signatures of Emergent Majorana Fermions and Fluxes, }}}{\bibfield  {journal} {\bibinfo  {journal} {Phys. Rev. Lett.}\ }\textbf
  {\bibinfo {volume} {112}},\ \bibinfo {pages} {207203} (\bibinfo {year}
  {2014})}\BibitemShut {NoStop}%
\bibitem [{\citenamefont {Vidal}(2007)}]{vidal_classical_2007}%
  \BibitemOpen
  \bibfield  {author} {\bibinfo {author} {\bibfnamefont {G.}~\bibnamefont
  {Vidal}},\ }\href {\doibase 10.1103/PhysRevLett.98.070201} {\bibfield {title} {\bibinfo {title} {Classical Simulation of Infinite-Size Quantum Lattice Systems in One Spatial Dimension, }}}{\bibfield
  {journal} {\bibinfo  {journal} {Phys. Rev. Lett.}\ }\textbf {\bibinfo
  {volume} {98}},\ \bibinfo {pages} {070201} (\bibinfo {year}
  {2007})}\BibitemShut {NoStop}%
\bibitem [{\citenamefont {Kimchi}\ \emph {et~al.}(2014)\citenamefont {Kimchi},
  \citenamefont {Analytis},\ and\ \citenamefont
  {Vishwanath}}]{kimchi_three-dimensional_2014}%
  \BibitemOpen
  \bibfield  {author} {\bibinfo {author} {\bibfnamefont {I.}~\bibnamefont
  {Kimchi}}, \bibinfo {author} {\bibfnamefont {J.~G.}\ \bibnamefont
  {Analytis}}, \ and\ \bibinfo {author} {\bibfnamefont {A.}~\bibnamefont
  {Vishwanath}},\ }\href {\doibase 10.1103/PhysRevB.90.205126} {\bibfield {title} {\bibinfo {title} {Three-dimensional quantum spin liquids in models of harmonic-honeycomb iridates and phase diagram in an infinite-D approximation, }}}{\bibfield
  {journal} {\bibinfo  {journal} {Phys. Rev. B}\ }\textbf {\bibinfo {volume}
  {90}},\ \bibinfo {pages} {205126} (\bibinfo {year} {2014})}\BibitemShut
  {NoStop}%
\bibitem [{\citenamefont {Banerjee}\ \emph {et~al.}(2016)\citenamefont
  {Banerjee}, \citenamefont {Yan}, \citenamefont {Knolle}, \citenamefont
  {Bridges}, \citenamefont {Stone}, \citenamefont {Lumsden}, \citenamefont
  {Mandrus}, \citenamefont {Tennant}, \citenamefont {Moessner},\ and\
  \citenamefont {Nagler}}]{Banerjee2016}%
  \BibitemOpen
  \bibfield  {author} {\bibinfo {author} {\bibfnamefont {A.}~\bibnamefont
  {Banerjee}}, \bibinfo {author} {\bibfnamefont {J.}~\bibnamefont {Yan}},
  \bibinfo {author} {\bibfnamefont {J.}~\bibnamefont {Knolle}}, \bibinfo
  {author} {\bibfnamefont {C.~A.}\ \bibnamefont {Bridges}}, \bibinfo {author}
  {\bibfnamefont {M.~B.}\ \bibnamefont {Stone}}, \bibinfo {author}
  {\bibfnamefont {M.~D.}\ \bibnamefont {Lumsden}}, \bibinfo {author}
  {\bibfnamefont {D.~G.}\ \bibnamefont {Mandrus}}, \bibinfo {author}
  {\bibfnamefont {D.~A.}\ \bibnamefont {Tennant}}, \bibinfo {author}
  {\bibfnamefont {R.}~\bibnamefont {Moessner}}, \ and\ \bibinfo {author}
  {\bibfnamefont {S.~E.}\ \bibnamefont {Nagler}},\ }\href
  {} {\bibfield {title} {\bibinfo {title} {Neutron scattering in the proximate quantum spin liquid $\alpha$-RuCl$_3$, }}}{\bibfield  {journal} {\bibinfo  {journal}
  {Science}\ }\textbf {\bibinfo {volume}
  {356}},\ \bibinfo {pages} {6342} (\bibinfo {year} {2017})}\BibitemShut
  {NoStop}%
\bibitem [{\citenamefont {Chaloupka}\ \emph {et~al.}(2010)\citenamefont
  {Chaloupka}, \citenamefont {Jackeli},\ and\ \citenamefont
  {Khaliullin}}]{Chaloupka2010}%
  \BibitemOpen
  \bibfield  {author} {\bibinfo {author} {\bibfnamefont {J.}~\bibnamefont
  {Chaloupka}}, \bibinfo {author} {\bibfnamefont {G.}~\bibnamefont {Jackeli}},
  \ and\ \bibinfo {author} {\bibfnamefont {G.}~\bibnamefont {Khaliullin}},\
  }\href@noop {} {\bibfield {title} {\bibinfo {title} {Kitaev-Heisenberg Model on a Honeycomb Lattice: Possible Exotic Phases in Iridium Oxides A$_2$IrO$_3$, }}}{\bibfield  {journal} {\bibinfo  {journal} {Phys. Rev. Lett.}\
  }\textbf {\bibinfo {volume} {105}},\ \bibinfo {pages} {027204} (\bibinfo
  {year} {2010})}\BibitemShut {NoStop}%
\bibitem [{\citenamefont {Kawamura}\ \emph {et~al.}(2017)\citenamefont
  {Kawamura}, \citenamefont {Yoshimi}, \citenamefont {Misawa}, \citenamefont
  {Yamaji}, \citenamefont {Todo},\ and\ \citenamefont
  {Kawashima}}]{Kawamura2017180}%
  \BibitemOpen
  \bibfield  {author} {\bibinfo {author} {\bibfnamefont {M.}~\bibnamefont
  {Kawamura}}, \bibinfo {author} {\bibfnamefont {K.}~\bibnamefont {Yoshimi}},
  \bibinfo {author} {\bibfnamefont {T.}~\bibnamefont {Misawa}}, \bibinfo
  {author} {\bibfnamefont {Y.}~\bibnamefont {Yamaji}}, \bibinfo {author}
  {\bibfnamefont {S.}~\bibnamefont {Todo}}, \ and\ \bibinfo {author}
  {\bibfnamefont {N.}~\bibnamefont {Kawashima}},\ }\href@noop {} {\bibfield {title} {\bibinfo {title} {Quantum lattice model solver $\mathcal{H}\Phi$, }}}{\bibfield
  {journal} {\bibinfo  {journal} {Computer Physics Communications}\ }\textbf
  {\bibinfo {volume} {217}},\ \bibinfo {pages} {180 } (\bibinfo {year}
  {2017})}\BibitemShut {NoStop}%
\end{thebibliography}

\begin{thebibliography}{12}%
\makeatletter
\providecommand \@ifxundefined [1]{%
 \@ifx{#1\undefined}
}%
\providecommand \@ifnum [1]{%
 \ifnum #1\expandafter \@firstoftwo
 \else \expandafter \@secondoftwo
 \fi
}%
\providecommand \@ifx [1]{%
 \ifx #1\expandafter \@firstoftwo
 \else \expandafter \@secondoftwo
 \fi
}%
\providecommand \natexlab [1]{#1}%
\providecommand \enquote  [1]{``#1''}%
\providecommand \bibnamefont  [1]{#1}%
\providecommand \bibfnamefont [1]{#1}%
\providecommand \citenamefont [1]{#1}%
\providecommand \href@noop [0]{\@secondoftwo}%
\providecommand \href [0]{\begingroup \@sanitize@url \@href}%
\providecommand \@href[1]{\@@startlink{#1}\@@href}%
\providecommand \@@href[1]{\endgroup#1\@@endlink}%
\providecommand \@sanitize@url [0]{\catcode `\\12\catcode `\$12\catcode
  `\&12\catcode `\#12\catcode `\^12\catcode `\_12\catcode `\%12\relax}%
\providecommand \@@startlink[1]{}%
\providecommand \@@endlink[0]{}%
\providecommand \url  [0]{\begingroup\@sanitize@url \@url }%
\providecommand \@url [1]{\endgroup\@href {#1}{\urlprefix }}%
\providecommand \urlprefix  [0]{URL }%
\providecommand \Eprint [0]{\href }%
\providecommand \doibase [0]{http://dx.doi.org/}%
\providecommand \selectlanguage [0]{\@gobble}%
\providecommand \bibinfo  [0]{\@secondoftwo}%
\providecommand \bibfield  [0]{\@secondoftwo}%
\providecommand \translation [1]{[#1]}%
\providecommand \BibitemOpen [0]{}%
\providecommand \bibitemStop [0]{}%
\providecommand \bibitemNoStop [0]{.\EOS\space}%
\providecommand \EOS [0]{\spacefactor3000\relax}%
\providecommand \BibitemShut  [1]{\csname bibitem#1\endcsname}%
\let\auto@bib@innerbib\@empty
\bibitem [{\citenamefont {Gohlke2}(2017)}]{Gohlke22017}%
  \BibitemOpen
  \bibfield  {author} {\bibinfo {author} {\bibfnamefont {M.}\ \bibnamefont
      {Gohlke}},\ \bibinfo {author} {\bibfnamefont {G.}\ \bibnamefont {Wachtel}},\
    \bibinfo {author} {\bibfnamefont {Y.}\ \bibnamefont {Yamaji}},\
    \bibinfo {author} {\bibfnamefont {F.}\ \bibnamefont {Pollmann}},\
    \bibinfo {author} {\bibfnamefont {Y. B.}\ \bibnamefont {Kim}},\    
  }\href {https://arxiv.org/abs/1706.09908} {\bibfield  {journal} {\bibinfo
  {journal} {arXiv preprint arXiv:1706.09908}\ } (\bibinfo {year}
  {2017})}\BibitemShut {NoStop}%
\bibitem [{\citenamefont {Albuquerque}\ \emph {et~al.}(2010)\citenamefont
  {Albuquerque}, \citenamefont {Alet}, \citenamefont {Sire},\ and\
  \citenamefont {Capponi}}]{Albuquerque2010}%
  \BibitemOpen
  \bibfield  {author} {\bibinfo {author} {\bibfnamefont {A.~F.}\ \bibnamefont
  {Albuquerque}}, \bibinfo {author} {\bibfnamefont {F.}~\bibnamefont {Alet}},
  \bibinfo {author} {\bibfnamefont {C.}~\bibnamefont {Sire}}, \ and\ \bibinfo
  {author} {\bibfnamefont {S.}~\bibnamefont {Capponi}},\ }\href@noop {}
  {\bibfield  {journal} {\bibinfo  {journal} {Phys. Rev. B}\ }\textbf {\bibinfo
  {volume} {81}},\ \bibinfo {pages} {064418} (\bibinfo {year}
  {2010})}\BibitemShut {NoStop}%
\bibitem [{\citenamefont {Reimann}(2007)}]{PhysRevLett.99.160404}%
  \BibitemOpen
  \bibfield  {author} {\bibinfo {author} {\bibfnamefont {P.}~\bibnamefont
  {Reimann}},\ }\href@noop {} {\bibfield  {journal} {\bibinfo  {journal} {Phys.
  Rev. Lett.}\ }\textbf {\bibinfo {volume} {99}},\ \bibinfo {pages} {160404}
  (\bibinfo {year} {2007})}\BibitemShut {NoStop}%
\bibitem [{\citenamefont {Imada}\ and\ \citenamefont
  {Takahashi}(1986)}]{JPSJ.55.3354}%
  \BibitemOpen
  \bibfield  {author} {\bibinfo {author} {\bibfnamefont {M.}~\bibnamefont
  {Imada}}\ and\ \bibinfo {author} {\bibfnamefont {M.}~\bibnamefont
  {Takahashi}},\ }\href {\doibase 10.1143/JPSJ.55.3354} {\bibfield  {journal}
  {\bibinfo  {journal} {J. Phys. Soc. Jpn.}\ }\textbf {\bibinfo {volume}
  {55}},\ \bibinfo {pages} {3354} (\bibinfo {year} {1986})}\BibitemShut
  {NoStop}%
\bibitem [{\citenamefont {Jaklic}\ and\ \citenamefont
  {Prelovsek}(1994)}]{PhysRevB.49.5065}%
  \BibitemOpen
  \bibfield  {author} {\bibinfo {author} {\bibfnamefont {J.}~\bibnamefont
  {Jaklic}}\ and\ \bibinfo {author} {\bibfnamefont {P.}~\bibnamefont
  {Prelovsek}},\ }\href@noop {} {\bibfield  {journal} {\bibinfo  {journal}
  {Phys. Rev. B}\ }\textbf {\bibinfo {volume} {49}},\ \bibinfo {pages} {5065}
  (\bibinfo {year} {1994})}\BibitemShut {NoStop}%
\bibitem [{\citenamefont {Hams}\ and\ \citenamefont
  {De~Raedt}(2000)}]{PhysRevE.62.4365}%
  \BibitemOpen
  \bibfield  {author} {\bibinfo {author} {\bibfnamefont {A.}~\bibnamefont
  {Hams}}\ and\ \bibinfo {author} {\bibfnamefont {H.}~\bibnamefont
  {De~Raedt}},\ }\href@noop {} {\bibfield  {journal} {\bibinfo  {journal}
  {Phys. Rev. E}\ }\textbf {\bibinfo {volume} {62}},\ \bibinfo {pages} {4365}
  (\bibinfo {year} {2000})}\BibitemShut {NoStop}%
\bibitem [{\citenamefont {Sugiura}\ and\ \citenamefont
  {Shimizu}(2012)}]{SSugiura2012}%
  \BibitemOpen
  \bibfield  {author} {\bibinfo {author} {\bibfnamefont {S.}~\bibnamefont
  {Sugiura}}\ and\ \bibinfo {author} {\bibfnamefont {A.}~\bibnamefont
  {Shimizu}},\ }\href@noop {} {\bibfield  {journal} {\bibinfo  {journal} {Phys.
  Rev. Lett.}\ }\textbf {\bibinfo {volume} {108}},\ \bibinfo {pages} {240401}
  (\bibinfo {year} {2012})}\BibitemShut {NoStop}%
\bibitem [{\citenamefont {Sugiura}\ and\ \citenamefont
  {Shimizu}(2013)}]{SSugiura2013}%
  \BibitemOpen
  \bibfield  {author} {\bibinfo {author} {\bibfnamefont {S.}~\bibnamefont
  {Sugiura}}\ and\ \bibinfo {author} {\bibfnamefont {A.}~\bibnamefont
  {Shimizu}},\ }\href@noop {} {\bibfield  {journal} {\bibinfo  {journal} {Phys.
  Rev. Lett.}\ }\textbf {\bibinfo {volume} {111}},\ \bibinfo {pages} {010401}
  (\bibinfo {year} {2013})}\BibitemShut {NoStop}%
\bibitem [{\citenamefont {Vidal}(2007)}]{vidal_classical_2007}%
  \BibitemOpen
  \bibfield  {author} {\bibinfo {author} {\bibfnamefont {G.}~\bibnamefont
  {Vidal}},\ }\href {\doibase 10.1103/PhysRevLett.98.070201} {\bibfield
  {journal} {\bibinfo  {journal} {Phys. Rev. Lett.}\ }\textbf {\bibinfo
  {volume} {98}},\ \bibinfo {pages} {070201} (\bibinfo {year}
  {2007})}\BibitemShut {NoStop}%
\bibitem [{\citenamefont {Sornborger}\ and\ \citenamefont
  {Stewart}(1998)}]{sornborger_higher-order_1998}%
  \BibitemOpen
  \bibfield  {author} {\bibinfo {author} {\bibfnamefont {A.~T.}\ \bibnamefont
  {Sornborger}}\ and\ \bibinfo {author} {\bibfnamefont {E.~D.}\ \bibnamefont
  {Stewart}},\ }\href {http://arxiv.org/abs/quant-ph/9809009} {\bibfield
  {journal} {\bibinfo  {journal} {arXiv:quant-ph/9809009}\ } (\bibinfo {year}
  {1998})},\ \bibinfo {note} {arXiv: quant-ph/9809009}\BibitemShut {NoStop}%
\bibitem [{\citenamefont {Or\'us}\ and\ \citenamefont
  {Vidal}(2008)}]{orus_infinite_2008}%
  \BibitemOpen
  \bibfield  {author} {\bibinfo {author} {\bibfnamefont {R.}~\bibnamefont
  {Or\'us}}\ and\ \bibinfo {author} {\bibfnamefont {G.}~\bibnamefont {Vidal}},\
  }\href {\doibase 10.1103/PhysRevB.78.155117} {\bibfield  {journal} {\bibinfo
  {journal} {Phys. Rev. B}\ }\textbf {\bibinfo {volume} {78}},\ \bibinfo
  {pages} {155117} (\bibinfo {year} {2008})}\BibitemShut {NoStop}%
\bibitem [{\citenamefont {Kitaev}(2006)}]{kitaev_anyons_2006}%
  \BibitemOpen
  \bibfield  {author} {\bibinfo {author} {\bibfnamefont {A.}~\bibnamefont
  {Kitaev}},\ }\href {\doibase 10.1016/j.aop.2005.10.005} {\bibfield  {journal}
  {\bibinfo  {journal} {Annals of Physics}\ }\bibinfo {series} {January
  {Special} {Issue}},\ \textbf {\bibinfo {volume} {321}},\ \bibinfo {pages} {2}
  (\bibinfo {year} {2006})}\BibitemShut {NoStop}%
\bibitem [{\citenamefont {Kimchi}\ \emph {et~al.}(2014)\citenamefont {Kimchi},
  \citenamefont {Analytis},\ and\ \citenamefont
  {Vishwanath}}]{kimchi_three-dimensional_2014}%
  \BibitemOpen
  \bibfield  {author} {\bibinfo {author} {\bibfnamefont {I.}~\bibnamefont
  {Kimchi}}, \bibinfo {author} {\bibfnamefont {J.~G.}\ \bibnamefont
  {Analytis}}, \ and\ \bibinfo {author} {\bibfnamefont {A.}~\bibnamefont
  {Vishwanath}},\ }\href {\doibase 10.1103/PhysRevB.90.205126} {\bibfield
  {journal} {\bibinfo  {journal} {Phys. Rev. B}\ }\textbf {\bibinfo {volume}
  {90}},\ \bibinfo {pages} {205126} (\bibinfo {year} {2014})}\BibitemShut
  {NoStop}%
\end{thebibliography}

%

\end{document}